\documentclass[superscriptaddress,aps,pre,nolongbibliography,twocolumn,showpacs,floatfix]{revtex4-1} 
\usepackage{graphicx}
\usepackage{amsmath}
\usepackage{amsfonts}
\usepackage{amssymb}
\usepackage{MnSymbol}

\def\Ut{\hat{U}_t}
\def\At{\hat{A}_t}
\def\A{\hat{A}}
\def\B{\hat{B}}
\def\O{\hat{O}}
\def\Ot{\hat{O}_t}
\def\H{\hat{H}}
\def\X{\hat{X}}
\def\P{\hat{P}}
\def\rv{\mathbf{r}}
\def\rvp{\mathbf{r}^{\prime}}
\def\rb{\mathbf{\bar r}}
\def\rbp{\mathbf{\bar r}^{\prime}}
\def\deltarv{\mathbf{\Delta r}}
\def\deltarvp{\mathbf{\Delta r}^{\prime}}
\def\hatrv{\mathbf{\hat r}}
\def\deltahatrv{\mathbf{\Delta {\hat r}}}
\def\deltahatrt{\Delta {\hat r}^{\bot}}
\def\deltahatrp{\Delta {\hat r}^{\parallel}}
\def\pv{\mathbf{p}}
\def\pvp{\mathbf{p}^{\prime}}
\def\pb{\mathbf{\bar p}}
\def\pbp{\mathbf{\bar p}^{\prime}}
\def\hatpv{\mathbf{\hat p}}
\def\deltahatpv{\mathbf{\Delta {\hat p}}}
\def\deltahatpt{\Delta {\hat p}^{\bot}}
\def\deltahatpp{\Delta {\hat p}^{\parallel}}
\def\uv{\mathbf{u}}
\def\vv{\mathbf{v}}
\def\ve{\varepsilon}
\def\d{{\rm d}}
\newcommand{\im}{{\rm i}}
\def\hatev{\mathbf{\hat e}}
\def\kb{k_{\rm B}} 

\begin{document}

\title{Semiclassical theory of out-of-time-order correlators for low-dimensional classically chaotic systems}

\author{Rodolfo A. Jalabert$^{1}$, Ignacio Garc\'{\i }a-Mata$^{2,3}$, and Diego A. Wisniacki$^{4}$\\
$^{1}$\textit{Universit\'e de Strasbourg, CNRS, Institut de Physique et Chimie des Mat\'eriaux de Strasbourg, UMR 7504, F-67000 Strasbourg, France}\\
$^{2}$\textit{Instituto de Investigaciones F\'{\i }sicas de Mar del Plata (IFIMAR), CONICET-UNMdP, Mar del Plata, Argentina}\\
$^{3}$\textit{Consejo Nacional de Investigaciones Cient\'{\i }ficas y Tecnol\'ogicas (CONICET), Argentina}\\
$^{4}$\textit{Departamento de F\'{\i }sica “J.J. Giambiagi” and IFIBA, FCEyN, Universidad de Buenos Aires, 1428 Buenos Aires, Argentina}
}
\begin{abstract}
The out-of-time-order correlator (OTOC), recently analyzed in several physical contexts, is studied for low-dimensional chaotic systems through semiclassical expansions and numerical simulations. The semiclassical expansion for the OTOC yields a leading-order contribution in $\hbar^2$ that is exponentially increasing with time within an intermediate, temperature-dependent, time-window. The growth-rate in such a regime is governed by the Lyapunov exponent of the underlying classical system and scales with the square-root of the temperature.
\end{abstract}

\maketitle

\section{Introduction}
\label{sec:Intro}
The present interest in manipulating quantum many-body systems and understanding the flow of quantum information naturally leads to consider the behavior of the four-point out-of-time-order correlator (OTOC) \cite{swingle18}. In contrast with the time-ordered products of quantum operators, widely used in the perturbative treatments of quantum field theories, the OTOC allows to address the subtleties related with thermalization and time-evolution of complex systems. First introduced in the context of semiclassical approaches to superconductivity theory \cite{lo69}, the relevance of the OTOC in the physics of black holes and the conjectured bound on its growth rate \cite{mss16} has spurred a sustained activity in the last few years
\cite{sbsh16,afi16,cg16,kgp17,rgg17,hmy17,letal17,cdp17,gbswb17,k18,rgg18,cz18,rur18,gsjrw18,rfu18,hirschetal18}.

The complexity of a quantum system, arising from its many-body and/or from the chaotic nature of the underlying classical dynamics, has been claimed to be essential in rendering a generic behavior of the OTOC which is universal with respect to the choice made for the quantum operators \cite{mss16}. The OTOC has then been considered as a probe of quantum chaos, unveiling the signatures of classical chaos on a quantum system, like the level statistics \cite{bgs84,LesHou89,Gutz-book,ullmo16}, the Loschmidt echo \cite{jalabert2001,goussev12,philtrans},
or the eigenstate thermalization hypothesis \cite{srednicki94,alonso_jain97,deutsch18}. A two-point correlation function of Heisenberg operators at different times, formally related to the OTOC, has been used to characterize the stochasticity in complex quantum systems since the beginning of quantum chaos studies \cite{dima1,dima2}. 

While some of the key predictions about the OTOC address quantum systems with many degrees of freedom \cite{mss16}, and a considerable body of work has taken many-body spin systems as a paradigm \cite{kgp17,cdp17}, it is important to understand  the behavior of the OTOC in quantum systems with few degrees of freedom, where a good description of the classical and quantum dynamics is achievable. Among the latter, quantum maps \cite{rgg17,cdp17,cz18,gsjrw18,rfu18} and two-dimensional billiards \cite{hmy17,rgg18} have recently allowed to connect the time evolution of the OTOC with the features of the underlying classical dynamics. In particular, establishing the relevance of classical chaos.

Quantum maps are defined by an evolution operator, rather than a Hamiltonian, and thus, the thermal aspects of the OTOC cannot be put in evidence. The important outcomes of the OTOC studies in quantum maps concern their short time behavior, exhibiting an exponential growth determined by twice the Lyapunov exponent of the classical map \cite{rgg17,cz18}, as well as their long-time behavior, where the approach to saturation is governed by the largest non-trivial Ruelle-Pollicot resonance \cite{gsjrw18}. Chaos is thus embedded in the short and long-time behavior of the OTOC of a quantum map, much alike as in the dynamics of the classical one. Numerical work in low-dimensional billiards  \cite{hmy17} did not allow to identify a clear exponential growth of the OTOC governed by the Lyapunov exponent, nor to put in evidence a remarkable difference between the classically chaotic and integrable systems within the regime of initial growth. 

In this work we develop a semiclassical approach to the OTOC applicable to low-dimensional classically chaotic systems. We perform a systematic expansion in powers of $\hbar$, determining the leading classical behavior of the different components of the OTOC, and then their next-order contribution scaling with $\hbar^2$. We determine the growth-rate of the OTOC in the low-temperature limit, as well as the saturation behavior for long times. We confront our semiclassical results with numerical quantum calculations performed in a two-dimensional billiard, and make the connection with the predicted bounds on the OTOC growth rate.

The paper is organized as follows. In Sec.~\ref{sec:oitopce} we present the basic definitions of the OTOC for a one-particle systems in the canonical ensemble, while in Sec.~\ref{sec:SAOTOC} we introduce the semiclassical formalism for obtaining the OTOC as a sum of three components. In Sec.~\ref{sec:DSOTOCc} we identify the ensemble of classical trajectories relevant for the semiclassical calculation of these components, providing their leading-order (classical limit) in Sec.~\ref{sec:LOCOTOCC}, and confronting them to numerical calculations in Sec.~\ref{sec:QNCOTOCC}. Section~\ref{sec:OTOCSL} presents the semiclassical limit of the OTOC, while Sec.~\ref{sec:LowTBOTOC} discusses the resulting low-temperature and long-time limits. The comparison with quantum numerical calculations of the OTOC are presented in Sec.~\ref{sec:QNCOTOC}, and Sec.~\ref{sec:conclusions} sums up the main conclusions of this work. Appendices A and B describe alternative semiclassical schemes not included in the main text. 

\section{OTOC in the one-particle canonical ensemble}
\label{sec:oitopce}

We consider the out-of-time-order correlator defined by
\begin{equation}
{\cal C}(t) =  \left\langle\left\langle \left[  \At, \B \right]^{\dagger}  \left[  \At, \B \right] \right\rangle\right\rangle  \, .
\label{eq:OTOC}
\end{equation}
Where $\A$ and $\B$ are two operators to be specified, while $\At = \Ut^{\dagger}  \A \Ut$ follows from the time evolution of the former. We use the standard notation 
$\Ut=e^{-\im \H t/\hbar}$ for the evolution operator under the Hamiltonian $\H$ of the system. In this context, the double angular bracket stands for the thermal averaging which, in the one-body physics that we deal with, yields for an arbitrary operator  $\O$ 
\begin{equation}
\left\langle\left\langle \O \right\rangle\right\rangle =  \frac{1}{Z} \
{\rm Tr}\left\{ e^{-\beta \H} \O \right\}\, ,
\label{eq:therav}
\end{equation}
where $Z$ is the canonical partition function and $\beta=(\kb T)^{-1}$, with $T$ the temperature and $\kb$ the Boltzmann constant. 

When $\A$ and $\B$ are chosen to be unitary operators, we have
\begin{equation}
{\cal C}(t) =  2 \left( 1 - {\rm Re} \left\{ \left\langle\left\langle  \At^{\dagger}  \B^{\dagger}  \At \B  \right\rangle\right\rangle \right\} \right) \, .
\label{eq:OTOCUP}
\end{equation}
If, however,  $\A$ and $\B$ are Hermitian operators, the OTOC is given by
\begin{equation}
{\cal C}(t) = - \left\langle\left\langle \left[  \At, \B \right]^{2} \right\rangle\right\rangle = 
 - 2 \ {\rm Re} \left\{ {\cal O}^{(1)}(t) \right\}  + {\cal O}^{(2)}(t) + {\cal O}^{(3)}(t) \, .
\label{eq:OTOCHP}
\end{equation}
We note ${\cal O}^{(j)}(t) = \left\langle\left\langle \Ot^{(j)} \right\rangle\right\rangle$ the three components of the OTOC, obtained from
\begin{subequations}
\label{allO} 
\begin{equation}
\Ot^{(1)} =  \At  \B  \At \B  \ ,
\end{equation}
\begin{equation}
\Ot^{(2)} =  \At  \B^2  \At  \ ,
\end{equation}
\begin{equation}
\Ot^{(3)} =  \B  \At^2 \B  \ .
\end{equation}
\end{subequations}
When $\B =  \A$ and/or at infinite temperature we have $\Ot^{(2)} =  \Ot^{(3)}$.

We will focus on the case in which $\A$ and $\B$ are Hermitian operators, which is the one where the measurement interpretation of the OTOC is straightforward \cite{swingle18}. To make explicit some of our calculations, we will choose $\A = \X$ and $\B = \P_{X}$. That is,  the $X$ component of the position and momentum operators of a two-dimensional one-particle system, respectively. However, we will indicate the extent up to which our results are generic in terms of the chosen operators.  

Using the spectral decomposition of the Green function $G(\rv,\rvp;\ve)$ between 
$\rvp$ and $\rv$ at energy $\ve$, we can write
\begin{align}
{\cal O}^{(j)}(t) &= - \frac{1}{\pi Z} \int  \d \ve \ \d \rvp \ \d \rv \ e^{-\beta \ve} \ \nonumber \\
& \ \ \ \ {\rm Im} 
\left\{G(\rv,\rvp;\ve)\right\} \ O^{(\it j)}(\rvp,\rv;t) \, .
\label{eq:OTOCImGF}
\end{align}
 We have defined (for $j=1,2$, $3$) the matrix element
\begin{equation}
O^{(j)}(\rvp,\rv;t) = \langle \rvp \left| \Ot^{(j)} \right| \rv \rangle  \, ,
\label{eq:Oj}
\end{equation}
and we employed the Green function associated to the Schr\"odinger equation, that is, the Fourier transform of the propagator
\begin{equation}
K(\rvp,\rv;t) = \langle \rvp \left| \Ut \right| \rv \rangle  \, .
\label{eq:propagator}
\end{equation}
The energy integral in Eq.~\eqref{eq:OTOCImGF} runs over the whole spectrum, while the space integrals are taken over the area ${\cal A}$ of the system. The partition function for a spinless particle in a billiard of area ${\cal A}$ is $Z={\cal A}m/(2\pi \hbar^2 \beta)$.
\section{Semiclassical approach to the OTOC}
\label{sec:SAOTOC}

For the evaluation of $O^{(j)}(\rvp,\rv;t)$, the insertion  of a complete basis of the position operator within the products that define $\Ot^{(j)}$ gives rise to the propagator Eq.~\eqref{eq:propagator}. Throughout this work, we will make extensive use of the semiclassical approximation for the later, that in the two-dimensional case is given by the expansion \cite{LesHou89, Gutz-book}
\begin{align}
K_{\rm sc}(\rvp,\rv;t) & = \left(  \frac{1}{2\pi \im {\hbar}}\right) \sum_{s(\rv,\rvp;t)}
C_{s}^{1/2} \nonumber \\
&  \ \ \ \ \exp{\left[\tfrac{\im}{\hbar}R_{s}(\rvp,\rv;t)-\im \tfrac{\pi}{2}\mu_{s}\right]  } \, ,
\label{eq:SCpropagator}
\end{align}
as the sum over all the classical trajectories $s(\rv,\rvp;t)$ joining the points
$\rv$ and $\rvp$ in a time $t$. We note \mbox{$R_{s}(\rvp,\rv;t)=\int_{0}^{t} \d \tau \mathcal{L}$} the Hamilton principal function, obtained from the integral of the Lagrangian $\mathcal{L}$ along the classical path, and $\mu$ the Maslov index that counts the number of conjugate points. The prefactor $C_{s}=\left|  \det\mathcal{B}_{s}\right|$ accounting for the conservation of the classical probability, is expressed in terms of the initial and final position components $b$ and $a$ as %
$(\mathcal{B}_{s})_{ab}=-\partial^{2}R_{s}/\partial r_{a}^{\prime} \partial r_{b}$. 
The semiclassical approximation is applicable when the typical Hamilton principal function (or the action) of the relevant classical trajectories is much larger than $\hbar$. In the case of billiards on which we will focus in this work, the previous conditions translate into $kL \gg 1$, where $k$ is the magnitude of the wavevector and $L$ the minimal trajectory length.

The terms $O^{(1)}(\rvp,\rv;t)$ and $O^{(2)}(\rvp,\rv;t)$ are given as sums over products of four trajectories, while $O^{(3)}(\rvp,\rv;t)$ has a simpler expression, as it is obtained by a sum over pairs of trajectories. The semiclassical approximation for the later term is given by
\begin{widetext}
\begin{align}
O^{(3)}_{\rm sc}(\rvp,\rv;t)  &  = - \frac{1}{\left(2\pi \hbar\right)^2}  \int \d \rv_{1}
\sum_{s_{2}(\rv_{1},\rvp;t)} \ \sum_{s_{1}(\rv,\rv_{1};t)} 
C_{s_{2}}^{1/2} \ C_{s_{1}}^{1/2} \
\left\{P_{X,s_{2}}^{\rm f} \left( X_{1} \right)^{2} P_{X,s_{1}}^{\rm i}\right\} 
\nonumber \\ &
\exp{\left[  \frac{\im}{\hbar}\left(R_{s_{1}}(\rv_{1},\rv;t)-R_{s_{2}}(\rvp,\rv_{1};t)\right)  -\im \frac{\pi}{2}\left( \mu_{s_{1}}-\mu_{s_{2}}\right)  \right]  } \, ,
\label{eq:O3semi}
\end{align}
\end{widetext}
when as $\A$ and $\B$ are taken, respectively, as the $X$ component of the position and momentum operators. We note $X_1=\rv_1.\hat{\bf e}_X$ and $P_X=\pv.\hat{\bf e}_X$, with $\hat{\bf e}_X$ the unit vector in the $X$-direction, while the indices ${\rm i}$ and ${\rm f}$ refer, respectively, to the initial and final condition of the corresponding trajectory. The choice of $A$ and $B$ is not crucial at this stage, provided that they are local operators allowing to express, within a semiclassical formalism, their action at the beginning or the end of each trajectory, as indicated by the term inside the curly bracket in Eq.~\eqref{eq:O3semi}. 

\begin{widetext}
The semiclassical approximation for the term $O^{(2)}(\rvp,\rv;t)$ is given by
\begin{align}
O^{(2)}_{\rm sc}(\rvp,\rv;t)  &  = \frac{1}{\left(2\pi \hbar \right)^4}  
\int \d \rv_{3} \ \d \rv_{2} \ \d \rv_{1}
\sum_{s_{4}(\rv_{3},\rvp;t)} \ \sum_{s_{3}(\rv_{2},\rv_{3};t)} 
\sum_{s_{2}(\rv_{1},\rv_{2};t)} \ \sum_{s_{1}(\rv,\rv_{1};t)} 
\nonumber \\ &
C_{s_{4}}^{1/2} \ C_{s_{3}}^{1/2} \ C_{s_{2}}^{1/2} \ C_{s_{1}}^{1/2} \
\left\{X_{3} \left( P_{X,s_{3}}^{\rm i}\right)^2 X_{1}\right\} 
\nonumber \\ &
\exp{\left[  \frac{\im}{\hbar}\left(R_{s_{3}}(\rv_{3},\rv_{2};t)-R_{s_{4}}(\rvp,\rv_{3};t)\right)  - \im \frac{\pi}{2}\left( \mu_{s_{3}}-\mu_{s_{4}}\right)  \right]  } \nonumber \\ &
\exp{\left[  \frac{\im}{\hbar}\left(R_{s_{1}}(\rv_{1},\rv;t)-R_{s_{2}}(\rv_{2},\rv_{1};t)\right)  - \im \frac{\pi}{2}\left( \mu_{s_{1}}-\mu_{s_{2}}\right)  \right]  } 
\,  .
\label{eq:O2semi}
\end{align}
\end{widetext}
A similar semiclassical expression is obtained for $O^{(1)}(\rvp,\rv;t)$, with the only modification respect to \eqref{eq:O2semi} of using $X_{3} P_{X,s_{3}}^{\rm i} X_{1} P_{X,s_{1}}^{\rm i} $ within the curly bracket.

\section{Diagonal scheme for the OTOC components}
\label{sec:DSOTOCc}

Independent trajectories where the corresponding phases are unrelated average out their contribution upon the spatial integrations. Therefore, we will only keep in the sums terms in which these phases are related. Noting $\tilde{s}_{j}$ the time-reversal symmetric of the trajectory $s_{j}$, the most obvious connection is when $\tilde{s}_{2}$ remains near to $s_{1}$ [in the cases of $O^{(1)}(\rvp,\rv;t)$, $O^{(2)}(\rvp,\rv;t)$, and $O^{(3)}(\rvp,\rv;t)$] and  in addition $\tilde{s}_{4}$ is close to $s_{3}$ 
[only relevant in the cases of $O^{(1)}(\rvp,\rv;t)$ and $O^{(2)}(\rvp,\rv;t)$]. Such a pairing scheme, which we call {\it diagonal}, and note with the subscript ``${\rm d}$,'' is the simplest one, and it will then be first considered. We stress at this point that this diagonal scheme does not simply imply a strict diagonal approximation matching each trajectory with its time-reversal, as usually meant by this level of approximation \cite{jalabert2001,goussev12}, but it rather incorporates the contributions of the trajectories in the neighborhood of a given one. 

\begin{figure}
\centerline{\includegraphics[width=0.48\textwidth]{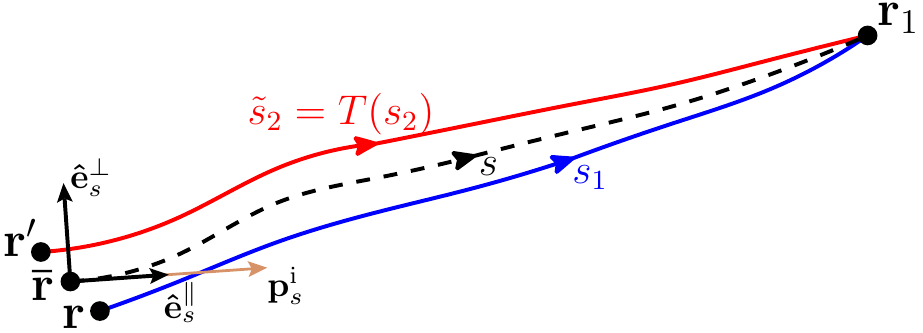}} 
\caption{Graphical representation of  $O^{(3)}(\rvp,\rv;t)$ according to Eq.~\eqref{eq:O3semi} for the case in which the trajectories $s_1$ and $s_2$ remain close to each other. We note ${\tilde s}_{2}=T(s_{2})$ the time reversed trajectory of $s_{2}$. The color blue (red) is used for trajectories whose Hamilton principal function appears with a plus (minus) sign in the phase term of Eq.~\eqref{eq:O3semi}. The dashed (black) line represents the trajectory $s$, leaving from $\rb=(\rv+\rvp)/2$ and reaching $\rv_{1}$ in a time $t$, used in Eq.~\eqref{eq:O3diag} to linearize the dynamics of nearby trajectories, while $\pv_{s}^{\rm i}$ stands for its initial momentum, and the unitary vectors $\hatev^{\parallel}_{s}$,  $\hatev^{\bot}_{s}$ define a local coordinate system. 
\label{fig:O3}}
\end{figure}

\begin{figure}
\centerline{\includegraphics[width=0.48\textwidth]{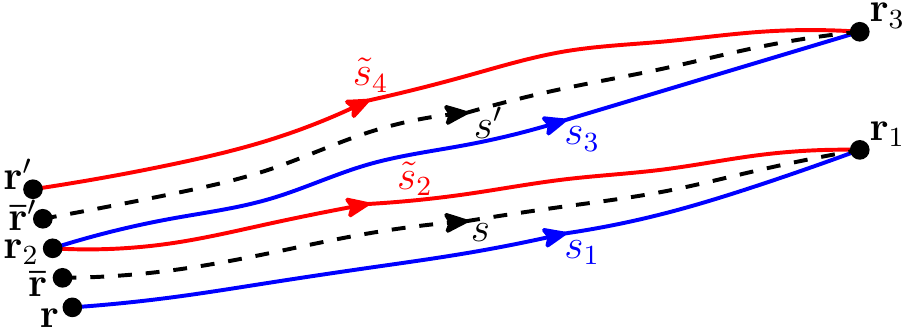}} 
\caption{
Graphical representation of $O^{(2)}(\rvp,\rv;t)$ according to Eq.~\eqref{eq:O2semi} for the case in which the trajectories $s_1$ and $s_2$, as well as $s_3$ and $s_4$, remain close to each other, using the same color convention as in Fig.~\ref{fig:O3}. The dashed (black) lines represent the trajectories $s$ ($s'$), leaving from $\rb=(\rv+\rv_{2})/2$ ($\rbp=(\rv_{2}+\rvp)/2$) and reaching $\rv_{1}$ ($\rv_{3}$) in a time $t$, used in Eq.~\eqref{eq:O2diag} to linearize the dynamics of nearby trajectories.
\label{fig:O2}}
\end{figure}

The graphical representation of Eqs.~\eqref{eq:O3semi} and \eqref{eq:O2semi} in the diagonal scheme is given in Figs.~\ref{fig:O3} and \ref{fig:O2}, respectively. The trajectories whose Hamilton principal function appears with a plus (minus) sign in the corresponding phase term  are indicated in blue (red). For the second case, the time-reversed trajectories ${\tilde s}_{j}=T(s_{j}) $ are considered, in view of their use in the forthcoming semiclassical calculation. Auxiliary trajectories, indicated in black, are taken as support of the linearization procedure. In  Appendix \ref{app:AP}, we consider an alternative pairing (Fig.~\ref{fig:O4}) with respect to the previous one, relevant for the cases of $O^{(1)}(\rvp,\rv;t)$ and $O^{(2)}(\rvp,\rv;t)$, showing that it does not result in an additional contribution within the leading order of the semiclassical calculation.

The semiclassical approach used for the propagator, together with the diagonal scheme, imply that the points $\rv$ and $\rvp$ of Eqs.~\eqref{eq:O3semi} and \eqref{eq:O2semi} remain close to each other. Thus, for the case of a billiard that we develop in this work, the Green function appearing in Eq.~\eqref{eq:OTOCImGF} can be expressed as the sum of the two-dimensional free-space Green function plus the contribution from nondirect classical trajectories joining $\rvp$ and $\rv$. The first term is
\begin{equation}
G_{2 {\rm D}}(\rv,\rvp;\ve) = - \frac{\im}{4} \ H_0\left(k|\rvp-\rv|\right) \  \frac{2m}{\hbar^2}   \, ,
\label{eq:freeGF}
\end{equation}
where $H_0$ is the zeroth-order Hankel function of the first kind, $m$ the particle mass, and $\ve=\hbar^2 k^2/2m$, while the second one has the form presented in Eq.~\eqref{eq:SCGF}. We first consider the contribution from Eq.~\eqref {eq:freeGF}, and in Appendix \ref{app:B} we provide the corrections arising from the nondirect trajectories. Since ${\rm Im} \{G_{2 {\rm D}}(\rv,\rvp;\ve)\} = -(1/4) J_0\left(k\left|\rvp-\rv\right|\right) 2m/\hbar^2$, with $J_0$ the zeroth-order Bessel function of the first kind, the $k$-integral stemming from Eq.~\eqref{eq:OTOCImGF} can be readily performed \cite{GR} and we obtain for the OTOC components in the diagonal scheme 
\begin{equation}
{\cal O}^{(j)}_{\rm d}(t) =  \frac{1}{\cal A} \int \d \rvp \ \d \rv \ \exp{\left[-\frac{m}{2 \beta \hbar^2} \left|\rvp-\rv\right|^2 \right]} \ O^{(j)}_{\rm d}(\rvp,\rv;t) \, .
\label{eq:OTOCint}
\end{equation}

For the spatial integrations of Eq.~\eqref{eq:OTOCint}, applied to ${\cal O}^{(3)}_{\rm d}(t)$, we perform a center-of-mass plus relative coordinate change from the original variables $(\rv,\rvp)$ to the new ones $(\rb,\deltarv)$ by posing $\rb=(\rv+\rvp)/2$ (see Fig.~\ref{fig:O3}) and $\deltarv=\rvp-\rv$. In the diagonal scheme, the trajectories $s_{1}$ and $s_{2}$ remain close to each other, therefore
%
\begin{align}
R_{s_{1}}(\rv_{1},\rv;t)-R_{s_{2}}(\rvp,\rv_{1};t) &= R_{s_{1}}(\rv_{1},\rv;t)-R_{{\tilde s}_{2}}(\rv_{1},\rvp;t) \nonumber \\
& \simeq \pv_{s}^{\rm i}.\deltarv \, 
\end{align}
up to quadratic order in $\deltarv$. The trajectory $s$ goes from $\rb$ to $\rv_{1}$ in time $t$, and we note $\pv_{s}^{\rm i}$ its initial momentum (see Fig.~\ref{fig:O3}). Within this approximation, ${\cal O}^{(3)}(t)$ reads
\begin{align}
{\cal O}^{(3)}_{\rm d}(t) &  = - \frac{1}{\left(2\pi \hbar\right)^2 {\cal A}}  \int \d \rb \ \d \deltarv \ \d \rv_{1} 
\ \exp{\left[-\frac{m}{2 \beta \hbar^2} \deltarv^2 \right]}  \nonumber\\
 & \sum_{s(\rb,\rv_{1};t)} C_{s} \
\left\{P_{X,s_{2}}^{\rm f} \left( X_{1} \right)^{2} P_{X,s_{1}}^{\rm i}\right\} 
\exp{\left[  \frac{\im}{\hbar} \pv_{s}^{\rm i}.\deltarv  \right]  } \, .
\label{eq:O3diag}
\end{align}
With the new integration variables, the trajectory $s_{1}({\tilde s}_{2})$ is the one that leaves $\rb \mp \deltarv/2$ and reaches $\rv_{1} $ in time $t$ while remaining in the environment of $s$. 

Making the change of variables from $\rv_{1}$ to $\pv_{s}^{\rm i}$ (that we note $\pb$) makes the prefactor $C_{s}$ (which is the Jacobian of the transformation) and the sum over $s$ to disappear from Eq.~\eqref{eq:O3diag}, leading to
\begin{align}
{\cal O}^{(3)}_{\rm d}(t) & = - \frac{1}{\left(2\pi \hbar\right)^2 \cal A}  \int \d \rb \ \d \deltarv \ \d \pb 
\ \nonumber \\
& \ \ \exp{\left[-\frac{m}{2 \beta \hbar^2} \deltarv^2 +
 \frac{\im}{\hbar} \pb.\deltarv \right]} \ 
\left\{P_{X,s_{2}}^{\rm f} \left( X_{1} \right)^{2} P_{X,s_{1}}^{\rm i}\right\}   \, ,
\label{eq:O3class0}
\end{align}
where the trajectory $s_{1} \left(s_{2}\right)$ joins the points $\rb \mp \deltarv/2$ and $\rv\left(\rb, \pb;t\right)$ in time $t$.

Similar lines as before can be applied for the calculation of ${\cal O}^{(2)}(t)$ from Eq.~\eqref{eq:OTOCint} within the diagonal scheme. However, in addition to the center-of-mass plus relative coordinate change from the variables $(\rv,\rv_{2})$ to $(\rb,\deltarv)$, we need to perform another one from the variables $(\rv_{2},\rvp)$ to $(\rbp,\deltarvp)$, yielding
\begin{widetext}
\begin{align}
{\cal O}^{(2)}_{\rm d}(t) &  = \frac{1}{\left(2\pi \hbar\right)^4 {\cal A}}  \int \d \rbp \ \d \deltarvp \ \d \rb \ \d \deltarv \  \d \rv_{3} \  \d \rv_{1} \
\delta \left(\rbp-\rb - \frac{\deltarvp + \deltarv}{2}\right) 
\ \exp{\left[-\frac{m}{2 \beta \hbar^2} \left(\rbp-\rb + \frac{\deltarvp + \deltarv}{2} \right)^2 \right]} 
\nonumber \\ &
\sum_{s^{\prime}(\rbp,\rv_{3};t)} \ \sum_{s(\rb,\rv_{1};t)}  C_{s^{\prime}} \ C_{s} 
\ \left\{X_{3} \left(P_{X,s_{3}}^{\rm i}  \right)^{2} X_{1}\right\} 
\ \exp{\left[  \frac{\im}{\hbar} \left( \pv_{s^{\prime}}^{\rm i}.\deltarvp + \pv_{s}^{\rm i}.\deltarv\right) \right]  }  \, .
\label{eq:O2diag}
\end{align}
With the new integration variables, the trajectory $s_{1}({\tilde s}_{2})$ is the one that leaves $\rb \mp \deltarv/2$ and reaches $\rv_{1} $ in time $t$ while remaining in the environment of $s$, and similarly $s_{3}({\tilde s}_{4})$ leaves $\rbp \mp \deltarvp/2$ and reaches $\rv_{3} $ in time $t$ while remaining in the environment of $s^{\prime}$ (see Fig.~\ref{fig:O2}). 

We now perform the changes of variable from $\rv_{1}$ to $\pv_{s}^{\rm i}$ (that we note $\pb$) and from $\rv_{3}$ to $\pv_{s^{\prime}}^{\rm i}$ (that we note $\pbp$), as well the center-of-mass plus relative coordinate change from the variables $(\rb,\rbp)$ to $(\hatrv,\deltahatrv)$. The integration over $\deltahatrv$ leads to 
\begin{align}
{\cal O}^{(2)}_{\rm d}(t) &  = \frac{1}{\left(2\pi \hbar\right)^4 {\cal A}}  \int \d \hatrv \ \d \deltarvp \ \d \deltarv \  \d \pb^{\prime} \  \d \pb \
\exp{\left[-\frac{m}{2 \beta \hbar^2} \left(\deltarvp + \deltarv \right)^2 +
\frac{\im}{\hbar} \left( \pbp.\deltarvp + \pb.\deltarv\right)
\right]} \  
\nonumber \\ &
\left\{X\left(\hatrv + \frac{\deltarvp + \deltarv}{4}, \pb^{\prime};t \right) \
\left(P_{X,s_{3}}^{\rm i}  \right)^{2} X\left(\hatrv - \frac{\deltarvp + \deltarv}{4},\pb;t \right)\right\}  \, ,
\label{eq:O2diag2}
\end{align}
where the trajectory $s_{3}$ joins the points $\hatrv - (\deltarvp - \deltarv)/4$ and $\rv\left(\hatrv + (\deltarvp + \deltarv)/4, \pb^{\prime};t\right)$ in time $t$.

Performing center-of-mass plus relative coordinate changes from the variables 
$(\deltarv,\deltarvp)$ to $(\uv,\vv)$ and from the variables $(\pb,\pbp)$ to $(\hatpv,\deltahatpv)$ we have
\begin{align}
{\cal O}^{(2)}_{\rm d}(t) &  = \frac{1}{\left(2\pi \hbar\right)^4 {\cal A}}  \int \d \hatrv \ \d \uv \ \d \vv \  \d \hatpv \  \d \deltahatpv \
\exp{\left[-\frac{2m}{\beta \hbar^2} \uv^2 +
\frac{\im}{\hbar} \left(2 \hatpv.\uv + \frac{\deltahatpv.\vv}{2} \right)
\right]} \ 
\nonumber \\ &
\left\{X\left(\hatrv + \frac{\uv}{2}, \hatpv + \frac{\deltahatpv}{2};t \right) \left(P_{X,s_{3}}^{\rm i}  \right)^{2} X\left(\hatrv - \frac{\uv}{2},\hatpv - \frac{\deltahatpv}{2};t \right)\right\} \, .
\label{eq:O2diag3}
\end{align}
\end{widetext}
With the new integration variables, the trajectory $s_{3}$ joins the points $\hatrv - \vv/4$ and $\rv\left(\hatrv + \uv/2,\hatpv + \deltahatpv/2;t\right)$ in time $t$.

Equations~\eqref{eq:O3class0} and \eqref{eq:O2diag3} are the basis of the systematic expansion of the OTOC components in powers of $\hbar$.
\section{Leading-order contribution of the OTOC components}
\label{sec:LOCOTOCC}

In the leading order of $\hbar$ for Eq.~\eqref{eq:O3class0}, we use a strict diagonal condition ${\tilde s}_{2}=s_{1}$ for the terms within the curly bracket, and take $P_{X,s_{2}}^{\rm f} = - P_{X,{\tilde s}_{2}}^{\rm i} = - P_{X,s_{1}}^{\rm i}$. We will call this approximation {\it classical}, and note the corresponding results by the index ``${\rm cl}$''. Writing $\deltarv = \deltahatrp \hatev^{\parallel}_{s} + \deltahatrt \hatev^{\bot}_{s}$, where $(\hatev^{\parallel}_{s},\hatev^{\bot}_{s})$ is the local coordinate system sketched in Fig.~\ref{fig:O3}, we perform the $\deltahatrp$ and $\deltahatrt$ integrals, which results in the purely classical expression
\begin{equation}
{\cal O}_{\rm cl}(t) = \frac{\beta}{2\pi {\cal A} m}  \int \d \rb  \ \d \pb 
\ \exp{\left[-\beta\frac{ \pb^2}{2 m} \right]} \ 
\{ {\bar P}_{X}^{2} \ X^{2}(\rb,\pb;t) \}  \, .
\label{eq:O3class}
\end{equation}
We have dropped the upper-index (3) from the last expression, since we will show in the sequel that the other components of the OTOC are also given by Eq.~\eqref{eq:O3class}. 

Turning to the calculation of ${\cal O}^{(2)}(t)$, to leading order in $\hbar$, we use the strict diagonal condition within the curly bracket of Eq.~\eqref{eq:O2diag3}. The $\vv$ and $\deltahatpv$-integrals can be trivially done, leading to 
\begin{align}
{\cal O}^{(2)}_{\rm cl}(t) &= \frac{1}{(\pi \hbar)^2{\cal A}}  \int \d \hatrv \ \d \uv \ \d \hatpv
\ \nonumber \\
& \ \ \exp{\left[-\frac{2m}{\beta \hbar^2} \uv^2 + \frac{2\im}{\hbar} \hatpv.\uv \right]} \ 
\{ {\hat P}_{X}^{2} \ X^{2}(\hatrv,\hatpv;t) \}  \, .
\label{eq:O2class}
\end{align}

The $\uv$-integration then yields the result Eq.~\eqref{eq:O3class}. At this level of approximation,  the calculation for ${\cal O}^{(1)}_{\rm cl}(t)$ is equivalent to that of ${\cal O}^{(2)}_{\rm cl}(t)$, and thus, also given by  Eq.~\eqref{eq:O3class}. 

We notice that the prefactor multiplying the integrals of  Eq.~\eqref{eq:O3class} can be simply cast as $[(2\pi \hbar)^2 Z]^{-1}$ and that the classical expression of the OTOC components directly follows from the thermal average of the operators explicit in  Eq.~\eqref{allO}, without the need of going through the previous semiclassical derivation. The usefulness of following the above-described procedure is to set the basis of a systematic semiclassical expansion for the OTOC. 

Since all components of the OTOC coincide in the classical limit, we necessarily have ${\cal C}_{\rm cl}(t)=0$. Such a result is not surprising, since the finite value of ${\cal C}(t)$ arises from noncommutation, which is a purely quantum concept. The classical limit of the OTOC components, even if canceling among themselves when evaluating ${\cal C}(t)$, are interesting to analyze, in particular in relation with their scaling with respect to the characteristic quantities of the problem. For instance, its zero-time value is simply given by
\begin{equation}
{\cal O}_{\rm cl}(t\!=\!0) = m \ {\cal G}_{Y}^{2} \ \kb T \, ,
\label{eq:teq0}
\end{equation}
where the gyration ratio ${\cal G}_{Y}$ of the stadium with respect to the $Y$-axis is defined by
\begin{equation}
{\cal G}_{Y}^{2} = \frac{1}{\cal A} \int \d \rv \ X^2  \, .
\end{equation}

For times smaller than the one corresponding to the first bounce off the billiard walls $\rv(\rb,\pb;t)=\rb+(\pb/m)t$, and then
\begin{equation}
{\cal O}_{\rm cl}(t) - {\cal O}_{\rm cl}(t\!=\!0) = 3 \left(\kb T \right)^2 t^2 = 3 m \ \kb T \ \ell^2 \, 
\label{eq:quadrbeh},
\end{equation}
where the length $\ell = {\tilde v} t$ provides the appropriate scaling. We use 
${\tilde v} = \left\langle V_X^2 \right\rangle^{1/2} = (\kb T/m)^{1/2}$ the root-mean-square for the $X$-component of the velocity in the free two-dimensional case. 

In a billiard, the energy of the trajectories is simply scalable, and therefore the momentum integral in Eq.~\eqref{eq:O3class} can be tackled by using polar coordinates. In following such a path, we see that ${\cal O}_{\rm cl}(t)$ scales linearly with the temperature $T$. The form of Eq.~\eqref{eq:O3class} applies even if we trade the operators ${\hat X}$ and ${\hat P}_{X}$  by arbitrary local ones ${\hat A}$ and ${\hat B}$. However, the resulting temperature scaling is characteristic of the particular choice of operators. 

\section{Quantum numerical calculation of the OTOC components}
\label{sec:QNCOTOCC}

\begin{figure}
\centerline{\includegraphics[width=0.48\textwidth]{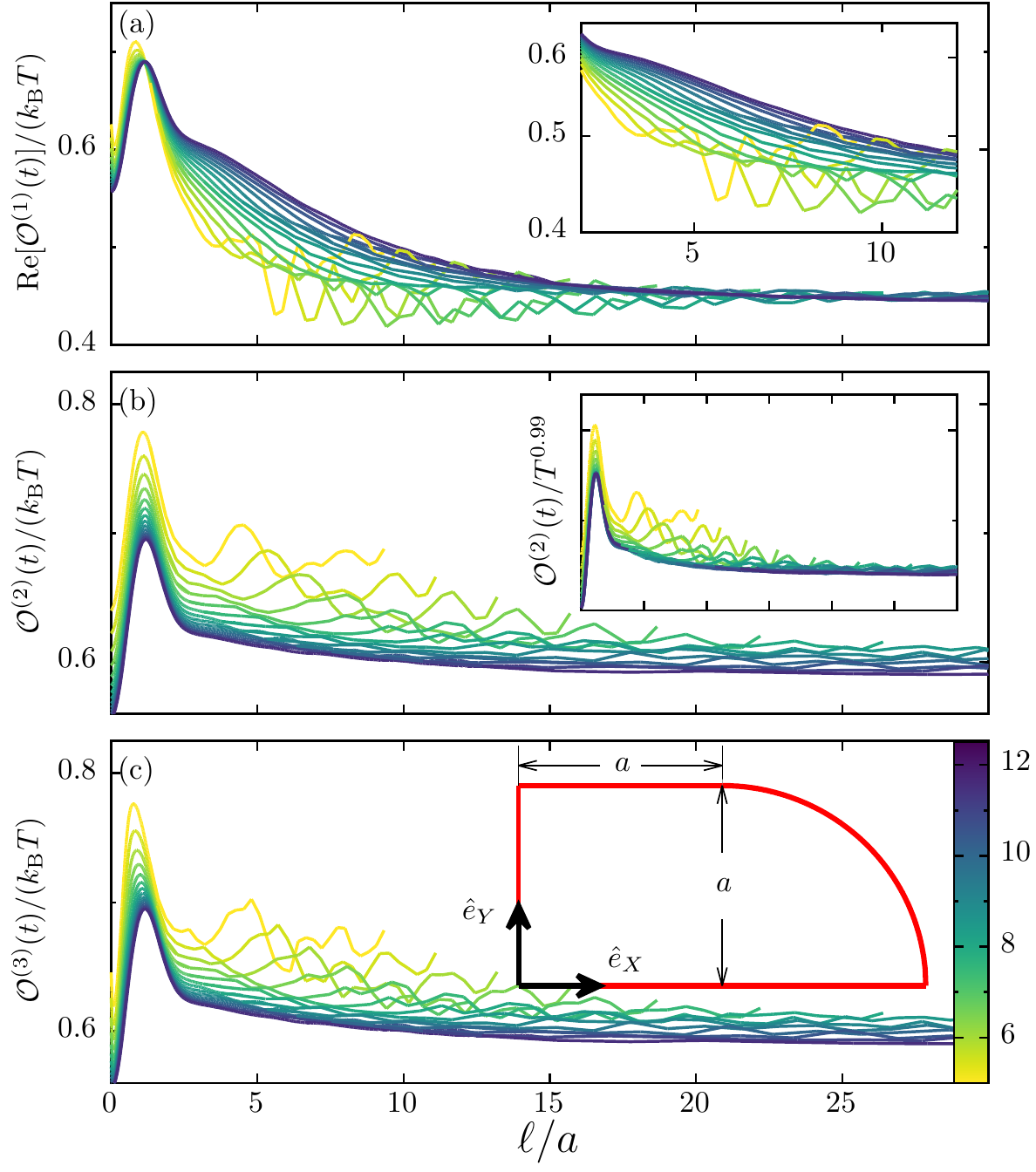}} 
\caption{
Numerically obtained OTOC components ${\cal O}^{(1)}(t)$ (a), ${\cal O}^{(2)}(t)$ (b), and ${\cal O}^{(3)}(t)$ (c) (divided by $\kb T$ and in units of $ma^2$) for the unsymmetrized stadium sketched in the inset of panel (c), as a function of the scaled time (length) $\ell = {\tilde v} t$ (in units of $a$), with $ {\tilde v} = (\beta m)^{-1/2}$ the mean-squared $X$-velocity component. The color code indicates the temperature scale, expressing $\kb T/E_0$ in a $\log_2$ basis, with $E_0=\hbar^2/(ma^2)$. The inset of panel (b) presents ${\cal O}^{(2)}(t)$, scaled with $T^{\xi}$ and $\xi=0.99$, in arbitrary units on the same $\ell/a$ interval of the main figure. The inset of panel (a) presents ${\cal O}^{(1)}(t)/(\kb T)$ in a logarithmic scale for the intermediate range of $\ell/a$ indicated in the horizontal axis. Within the leading order in $\hbar$, the quantum numerical results for ${\cal O}^{(j)}(t)$ should be compared with the classical expression Eq.~\eqref{eq:O3class}.
\label{fig:Oj}}
\end{figure}

Since ${\cal O}_{\rm cl}(t)$ is the leading contribution to all the ${\cal O}^{(j)}(t)$, it is meaningful to compare the quantum calculations of the latter with our theoretical prediction for the former, {\it i.e.} Eq.~\eqref{eq:O3class}. In Fig.~\ref{fig:Oj}, we show the quantum numerical calculations of ${\cal O}^{(j)}(t)$, at various temperatures $T$ (indicated by the color scale), for the unsymmetrical stadium sketched in the inset of Fig.~\ref{fig:Oj}(c).

For the numerical calculations of the OTOC and its components, we first compute the eigenstates of the stadium using the scaling method introduced in Ref.~\cite{vergini95}. We then build the matrix associated with the ${\hat X}$ operator in the eigenenergy basis, and subsequently the ${\hat P}_X$ matrix, using that for the stadium $[{\hat H},{\hat X}]=-2 \im {\hat P}_X$. Therefore the matrix elements of ${\hat P}_X$ can be obtained as
\begin{equation}
(P_X)_{mn}=\frac{\im}{2} \ E_{mn} \ X_{mn} \, ,
\end{equation}
where $E_{mn}=E_m-E_n$, $X_{mn}=\langle m | {\hat X} |n \rangle$, and
$(P_X)_{mn}=\langle m | {\hat P}_X |n \rangle$, with 
$|m\rangle$ ($|n\rangle$) the energy eigenstate corresponding to the energy $E_m (E_n)$. We truncate the basis to $5600$ states, and  we use the first $5000$. We have checked that for the temperatures that we show here, the results are converged as a function of the basis size. At the highest temperatures that we consider, the restriction of working with a finite basis starts to limit the accuracy of the numerical calculations. 

The scaling of the time (through the length $\ell = {\tilde v} t$) and of the values of the OTOC components (with the temperature $T$) are chosen in view of the theoretical predictions for ${\cal O}_{\rm cl}(t)$. While the linear-in-$T$ scaling of the OTOC components works quite well, we notice that a slightly better data collapsing of the numerical data is obtained when using a scaling of the form $T^{\xi}$ (with $\xi=1.03$ for ${\cal O}^{(1)}(t)$, and $\xi=0.99$ for ${\cal O}^{(2)}(t)$ and ${\cal O}^{(3)}(t)$). Such a scaling, applied to ${\cal O}^{(2)}(t)$ is presented as an inset in Fig.~\ref{fig:Oj}(b).  The smallness of the difference between the theoretical estimation for the leading-order contribution and the quantum mechanical calculation of the OTOC components makes it difficult to advance a possible explanation for such a discrepancy.

We can see that, for a given temperature and time, all three ${\cal O}^{(j)}(t)$ are of the same order, and thus ${\cal C}(t)$, that results from their difference, will in general be much smaller than each of them. The numerically obtained values of ${\cal O}^{(j)}(t\!=\!0)$, and the initial quadratic  increase with $\ell$, are in qualitative agreement with the classical results predicted by Eqs.~\eqref{eq:teq0} and \eqref{eq:quadrbeh}, respectively.  The oscillations as a function of $\ell$, visible at low temperatures, reflect the dynamics of the billiard and the signature of the periodic-orbit corrections (see Appendix \ref{app:B}). We remark that the mean-trajectory-length at temperature $T$ is $\sqrt{\pi/2} \ \ell$.  

According to Eq.~\eqref{eq:O3class}, the long-time saturation value for the OTOC components should be equal to ${\cal O}_{\rm cl}(t=0)$, due to the uniform distribution of $X(\rb,\pb;t)$ for $t \rightarrow \infty$. However, the numerical results yield a smaller saturation value for ${\cal O}^{(1)}(t)$, and slightly higher one for ${\cal O}^{(2)}(t)$ and ${\cal O}^{(3)}(t)$, pointing to the importance of periodic-orbit (discussed in Appendix \ref{app:B}) and other quantum corrections in the long-time limit. The exponential decrease of ${\cal O}^{(1)}(t)$ for $2 \lesssim \ell/a \lesssim 10$, before approaching a saturation value smaller than that of ${\cal O}^{(2)}(t)$ and ${\cal O}^{(3)}(t)$ (see inset of Fig.~\ref{fig:Oj}.a), is in line with the behavior obtained for unitary quantum maps \cite{gsjrw18}.
\section{OTOC in the semiclassical limit}
\label{sec:OTOCSL}

As discussed in the previous chapters, the OTOC is determined by the quantum corrections to its vanishing in the classical limit. Leaving aside the cases treated in the appendices, the quantum corrections stem from the difference between the diagonal scheme (presented in Sec.~\ref{sec:DSOTOCc}) and its classical limit (or strict diagonal approximation used in Sec.~\ref{sec:LOCOTOCC}). In order address such corrections, we  undertake the case of ${\cal O}^{(1)}_{\rm d}(t)$, for which the curly bracket of 
Eq.~\eqref{eq:O2diag} should be replaced by $X_{3} P_{X,s_{3}}^{\rm i} X_{1} P_{X,s_{1}}^{\rm i}$. Following the same procedure as with ${\cal O}^{(2)}_{\rm d}(t)$, instead of Eq.~\eqref{eq:O2diag3} we have
\begin{widetext}
\begin{align}
{\cal O}^{(1)}_{\rm d}(t) &  = - \frac{1}{\left(2\pi \hbar\right)^4 {\cal A}}  \int 
\d \hatrv \ \d \deltahatrv \ \d \uv \ \d \vv \  \d \hatpv \  \d \deltahatpv \
\delta \left(\deltahatrv - \uv \right) \
\exp{\left[-\frac{m}{2 \beta \hbar^2} \left(\deltahatrv + \uv \right)^2 \right]} \ 
\nonumber \\ &
\left\{X\left(\hatrv + \frac{\uv}{2}, \hatpv + \frac{\deltahatpv}{2};t \right) P_{X,s_{3}}^{\rm i}  X\left(\hatrv - \frac{\uv}{2},\hatpv - \frac{\deltahatpv}{2};t \right) P_{X,s_{1}} ^{\rm i} \right\} \
\exp{\left[  \frac{\im}{\hbar} \left(2 \hatpv.\uv + \frac{\deltahatpv.\vv}{2} \right) \right]  } \, .
\label{eq:O1diag3}
\end{align}
\end{widetext}
The trajectory $s_{1}$ joins the points $\hatrv -\uv + \vv/4$ and $\rv\left(\hatrv - \uv/2,\hatpv - \deltahatpv/2;t\right)$ in time $t$. The trajectory $s_{3}$, as a function of the new variables, has been defined after Eq.~\eqref{eq:O2diag3}.   

Within the diagonal scheme, but going beyond the classical approximation of Sec.~\ref{sec:LOCOTOCC}, we must incorporate the effect of having $s_{1} \neq {\tilde s}_{2}$ and $s_{3} \neq {\tilde s}_{4}$. Towards such a goal, we start by linearizing the dynamics of ${\tilde s}_{1}$ around that of ${\tilde s}$. Working with the time-reversal trajectories is not crucial, but has an easier visualization since these paths are diverging from a common initial point (see Figs.~\ref{fig:O3} and \ref{fig:O2}). Using local coordinates parallel and perpendicular to ${\tilde s}$, we can link the initial and final displacements with respect to this path through the corresponding stability matrix $S_{\tilde s}$
\begin{equation}
\left(
\begin{array}{c}
 \delta q^{\parallel }_{t} \\
  \delta q^{\bot }_{t} \\
  \delta p^{\parallel }_{t} \\
  \delta p^{\bot }_{t} \\
\end{array}
\right) = S_{\tilde s}
\left(
\begin{array}{c}
 \delta q^{\parallel }_{0} \\
  \delta q^{\bot }_{0} \\
  \delta p^{\parallel }_{0} \\
  \delta p^{\bot }_{0} \\
\end{array}
\right)
\end{equation}

Using the monodromy matrix of the dynamics on Riemann surfaces of constant negative curvature characterized by a Lyapunov exponent $\lambda$ , and the fact that $\delta q^{\parallel }_{0} =
  \delta q^{\bot }_{0} = 0$ for the displacement of ${\tilde s}_{1}$ with respect to ${\tilde s}$, we have
\begin{subequations}
\label{eq:alldis}
\begin{eqnarray} 
\rv-\rb &=& - \frac{t}{m} \ \delta p^{\parallel }_{0} \ \hatev^{\parallel }_{s} -
\frac{\sinh{(\lambda t)}}{m\lambda} \ \delta p^{\bot }_{0} \ \hatev^{\bot }_{s}
\, ,
\label{eq:disr}
\\ 
\pv_{s_1}^{\rm i}-\pb &=& \delta p^{\parallel }_{0} \ \hatev^{\parallel }_{s} +
\cosh{(\lambda }t) \ \delta p^{\bot }_{0} \ \hatev^{\bot }_{s}  \,  .
\label{eq:disp}
\end{eqnarray}
\end{subequations}  
In the left-hand-side we have used the initial displacement of $s_{1}$, which is the final one of ${\tilde s}_{1}$ (with the inversion of the momentum). In the right-hand-side we kept the intial displacement of the time reversed trajectory and used, as in Fig.~\ref{fig:O3}, the local unit vectors $\hatev^{\parallel}_{s}$ and $\hatev^{\bot}_{s}$ parallel and perpendicular, respectively, to $s$ at its initial point. The projection of the transverse momentum and position displacements of Eq.~\eqref{eq:alldis} on the $\hatev^{\bot}_{s}$ direction are related  through \cite{Goussev2008}
\begin{equation}
\left(\pv_{s_1}^{\rm i}-\pb\right)^{\bot} \simeq m\lambda \left( \rv-\rb \right)^{\bot} \, ,
\end{equation}
for long enough trajectories such that $\lambda t \gg 1$. And a similar relationship holds for $s_{3}$, which remains close to $s'$. We then have
\begin{align}
P_{X,s_{3}}^{\rm i}  P_{X,s_{1}}^{\rm i} & \simeq  {\hat P}_{X}^{2}  - \left(\frac{\deltahatpv_X}{2} \right)^2  + \left(\frac{ m\lambda}{2} \right)^2 \left(\left(\uv^{\bot }_{X} \right)^{2} - \left(\frac{\vv^{\bot }_{X}}{2}  \right)^{2} 
 \right) \nonumber \\ &
\ \ +m\lambda \left({\hat P}_{X}.\uv^{\bot }_{X}  - \frac{\deltahatpv_X.\vv^{\bot }_{X}}{8} \right) \, .
\label{eq:Pproduct}
\end{align} 

Along the same lines, we can linearize $s$ and $s'$ around the trajectory ${\hat s}$ that leaves from $\hatrv$ with momentum $\hatpv$ taking a time $t$. We then have
\begin{equation}
\rv_{3,1} = \rv\left(\hatrv,\hatpv;t \right) \pm \frac{1}{2} \left[ S_{\hat s}
\left(
\begin{array}{c}
 \deltahatrp \\
  \deltahatrt \\
  \deltahatpp \\
  \deltahatpt \\
\end{array}
\right)
\right]_{\rm rs} \, ,
\end{equation}
where the notation $[\ldots]_{\rm rs}$ indicates that only the real-space coordinates of the four-dimensional phase-space vector are considered.

For long enough trajectories, only the exponentially increasing terms of the stability matrix are kept, and we can furthermore assume that the direction of the final momentum is isotropically distributed. Therefore,  
\begin{align}
X\left(\rbp, \pbp;t \right)X\left(\rb, \pb;t \right) - X^2\left(\hatrv, \hatpv;t \right)  \simeq \ 
\nonumber\\
& \hspace{-2cm}-\frac{e^{2\lambda t}}{8} \left(\deltahatrt + 
\frac{ 1}{m \lambda} \deltahatpt 
 \right)^{2} \, .
\label{eq:Xproduct}
\end{align}
Within the previous approximation, the time-independent correction arising from Eq.~\eqref{eq:Pproduct} is irrelevant. Clearly, there appears a limitation to the validity of the exponential increase in the trajectory separation used in Eq.~\eqref{eq:Xproduct}, that cannot hold beyond the typical size of the system (of the order of $a$ for the billiard that we study). Such a restriction defines critical values of $\deltahatrt$ and $\deltahatpt$ beyond which the integrand is no longer exponentially increasing with time. In the next section, we discuss this issue, which is crucial to describe the long-time limit of the OTOC. Staying away from such a limit, we adopt Eq.~\eqref{eq:Xproduct}, and thus we write
\begin{widetext}
\begin{align}
\delta  {\cal O}^{(1)}(t) & = {\cal O}^{(1)}_{\rm d}(t) -  {\cal O}_{\rm cl}(t)  
= - \frac{1}{8 \left(2\pi \hbar\right)^4 {\cal A}}  
\int \d \hatrv \ \d \deltahatrv \ \d \uv \ \d \vv \  \d \hatpv \  \d \deltahatpv \
\delta \left(\deltahatrv - \uv \right)
\nonumber \\ &
\exp{\left[-\frac{m}{2 \beta \hbar^2} \left(\deltahatrv + \uv \right)^2 \right]} \ 
\left\{ e^{2\lambda t} \left(\deltahatrt + 
\frac{ 1}{m \lambda} \deltahatpt 
 \right)^{2} {\hat P}_{X}^{2} \right\} \
\exp{\left[  \frac{\im}{\hbar} \left(2 \hatpv.\uv + \frac{\deltahatpv.\vv}{2} \right) \right]  } \, .
\label{eq:deltaO1}
\end{align}

Performing the $\deltahatrv$, $\vv$, and $\deltahatpv$ integrals, we have
\begin{equation}
\delta  {\cal O}^{(1)}(t) = - \frac{1}{2(2 \pi \hbar)^2{\cal A} }  \int \d \hatrv \ \d \uv \ \d \hatpv
\ \exp{\left[-\frac{2m}{\beta \hbar^2} \uv^2 + \frac{2\im}{\hbar} \hatpv.\uv \right]} \ 
\left\{ e^{2\lambda t} \left(\uv^{\bot } \right)^{2}  
 {\hat P}_{X}^{2} \right\}  \, . 
\label{eq:deltaO1p}
\end{equation}

Using the decomposition $\uv= U^{\parallel } \ \hatev^{\parallel}_{\tilde s} + U^{\bot } \ \hatev^{\bot}_{\tilde s}$, the $U^{\parallel }$ and $U^{\bot }$ integrals can be readily done. At the level of approximation that we are working with $\delta  {\cal O}^{(2)}(t)=\delta  {\cal O}^{(1)}(t)$ and $\delta  {\cal O}^{(3)}(t)=0$, thus
\begin{equation}
{\cal C}(t) = - \delta  {\cal O}^{(1)}(t) = \frac{1}{64\pi}  \ 
\frac{\beta^2 \hbar^2}{{\cal A} m^2}  \int \d \hatrv  \ \d \hatpv 
\ \exp{\left[-\beta\frac{ \hatpv^2}{2 m} \right]} \ 
\left\{ e^{2\lambda\left(|\hatpv| \right) t} \ {\hat P}_{X}^{2} \right\}  \, .
\label{eq:Csemiclass}
\end{equation}
We have stressed the $|\hatpv|$ dependence of $\lambda$ in the last equation. The $\hatrv$-integral is now trivial and it simply leads to the cancellation of the factor ${\cal A}$. As expected, the OTOC scales with $\hbar^2$. The corresponding prefactor is a purely classical (time and temperature-dependent) one.
\end{widetext}
\section{Long-time and low-temperature  behavior of the OTOC}
\label{sec:LowTBOTOC}

In a billiard we have $\lambda\left(|\hatpv| \right) t = \lambda_{\rm g} L $, with $L = (|\hatpv|/m)t$ the trajectory length and $\lambda_{\rm g}= (m/|\hatpv|) \lambda$ a purely geometrical Lyapunov exponent. Using polar coordinates, the angular part of the $\hatpv$-integration can be readily done, leading to 
\begin{equation}
{\cal C}(t) =  \frac{\beta^2 \hbar^2}{64 m^2} 
\int \d p \ \ p^{3} \ \exp{\left[-\beta\frac{p^2}{2 m} + \frac{2\lambda_{\rm g} t}{m} p\right]} \ 
 \, ,
\label{eq:Csemiclass2}
\end{equation}
The $p$-integration admits a closed expression in terms of the error function. However, its capability to predict the short and long-time behavior of the OTOC is limited by the various approximations leading to Eq.~\eqref{eq:Csemiclass2}. In particular,  
the above-discussed limitations of Eq.~\eqref{eq:Xproduct} for describing large times or momenta. Such a finite-size effect can be accounted for by separating the $\uv$-integral of Eq.~\eqref{eq:deltaO1p} in  two pieces. The first one, up to $|\uv^{\bot}| \sim a \ e^{-2\lambda t}$, has the same integrand than Eq.~\eqref{eq:deltaO1p}, while for larger values of $|\uv^{\bot}|$ the curly bracket can be traded by $a^2 {\hat P}_{X}^{2}$. The second contribution dominates in the long-time limit, leading to a saturation value
\begin{equation}
{\cal C}_{\rm s} \varpropto m a^2  \kb T \, ,
\label{eq:CLongT}
\end{equation}
which scales with the temperature and the area of the billiard, but is independent of $\hbar$, in agreement with the findings of Ref.~\onlinecite{hmy17}. 

For sufficiently  low temperatures and not too long times, the term $e^{2\lambda_{\rm g} p t/m}$ in Eq.~\eqref{eq:Csemiclass2} behaves a smooth function of $p$ that can be taken outside of the integral, leading to a growth of ${\cal C}(t)$ with a rate $\Lambda$ given by the Lyapunov exponent corresponding to the average velocity. Such a procedure can be formalized by tackling the $p$-integral Eq.~\eqref{eq:Csemiclass} by the steepest-descent method, which for small $ \lambda_{\rm g} {\tilde v} t$ leads to
\begin{equation}
\frac{{\cal C}_{\rm LT}(t)}{\hbar^2} \varpropto 
\exp{\left[\sqrt{3} \ \lambda_{\rm g} {\tilde v} t \right]}  \, ,
\label{eq:CLowT}
\end{equation}
We then have $\Lambda=\sqrt{3} \ \lambda_{\rm g} {\tilde v}$ for an intermediate time-window and sufficiently low temperatures. For increasing temperatures, the previous reasoning stops being valid and there is no time-window where the exponential growth of the OTOC can be observed. 
 
\section{Quantum numerical calculation of the OTOC}
\label{sec:QNCOTOC}

\begin{figure}
\centerline{\includegraphics[width=0.48\textwidth]{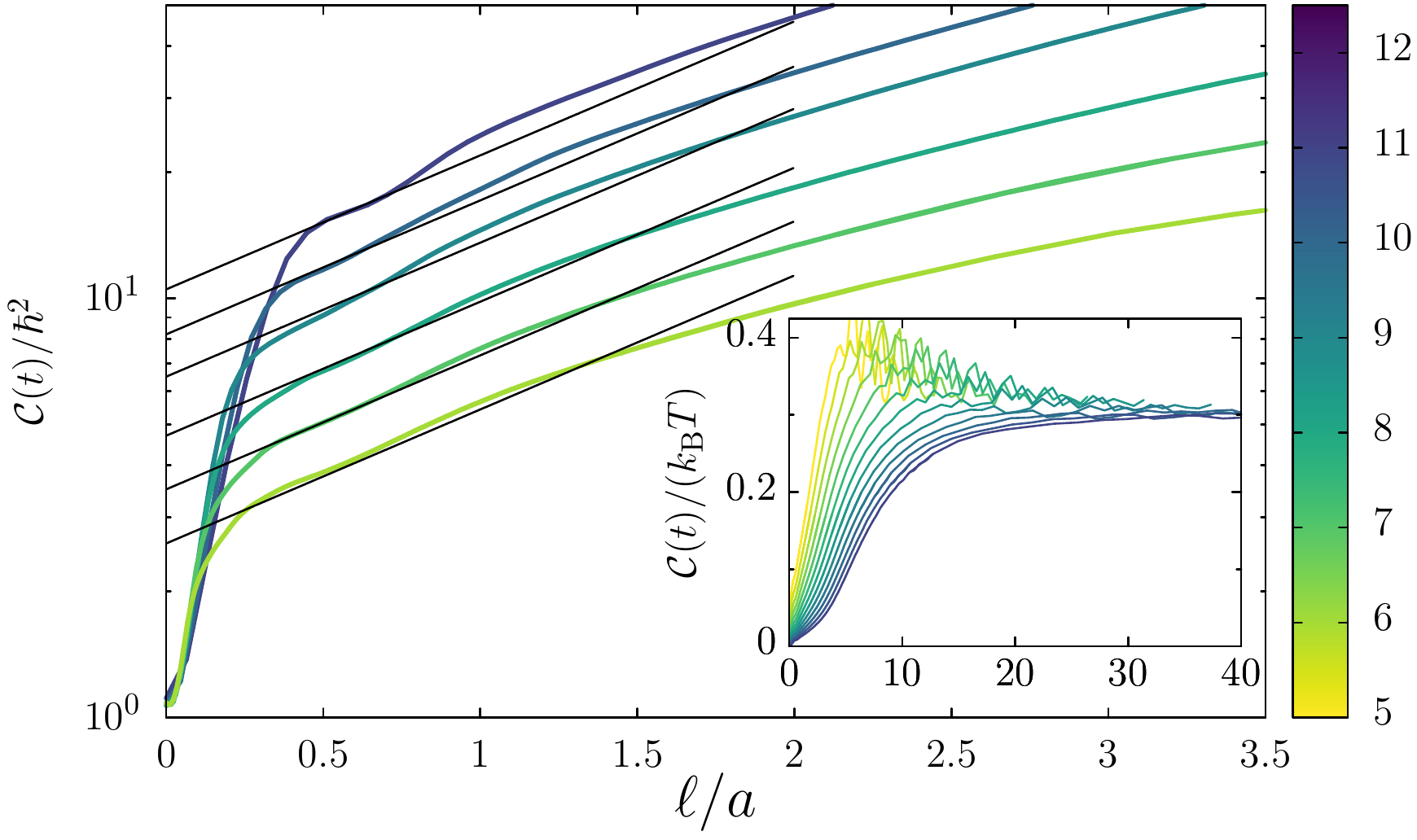}} 
\caption{
Numerically obtained OTOC (in a logarithmic scale) as a function of the scaled time (length) $\ell = {\tilde v} t$ for different temperatures according to the color scale (following the same conventions than in Fig.~\ref{fig:Oj}). The black straight lines describe the corresponding exponential growth $e^{\sqrt{3} \lambda_{\rm g} \ell}$ applicable in an intermediate time-window. Inset: OTOC scaled with the temperature (in a linear scale) in a larger $\ell/a$ interval showing its saturation in the long-time limit.
\label{fig:tse}}
\end{figure}

In Fig.~\ref{fig:tse} we present the quantum numerical calculations of the OTOC, according to the procedure described in Sec.~\ref{sec:QNCOTOCC}, for the billiard sketched in Fig.~\ref{fig:Oj}.c. For a given temperature and time, the typical values
of  ${\cal C}(t)$ are much smaller than those of its components ${\cal O}^{(j)}(t)$ (except for the long-time saturation values), confirming that the OTOC results from the small quantum correction of its components. 

For increasing values of $\ell$, we first notice a narrow regime with a quadratic take-off, followed by a regime of rapid growth. Further on, we identify a $\ell$-window with an exponential increase of the OTOC, well fitted when using the rate $\Lambda$ estimated in Eq.~\eqref{eq:CLowT} and the value $\lambda_g=0.425 a^{-1}$ applicable to the chosen billiard (black solid lines). The saturation of the OTOC for large times (see inset) completes the listing of observed regimes.  

The initial quadratic dependence (on $\ell$ or $t$) is characteristic of quantum perturbation theory, and it is thus expected in any kind of system. The ensuing rapid growth is also observed in a rectangle billiard (data not shown), as well as in other integrable systems (i.e. circle and particle-in-a-box) \citep{hmy17}. The semiclassical approach of  Secs.~\ref{sec:OTOCSL} and \ref{sec:LowTBOTOC} does not allow to access such a regime, corresponding to times much smaller than that of the first collision (which for a trajectory with the mean velocity occurs for $\ell/a \simeq 0.4$). The differences between chaotic and integrable systems appear in the regimes occurring for $\ell/a \gtrsim 0.4$. For the latter case, it is not possible to identify a window of exponential growth as in Fig.~\ref{fig:tse}, and also, no clear saturation of ${\cal C}(t)$ is observed.
 
\begin{figure}
\centerline{\includegraphics[width=0.48\textwidth]{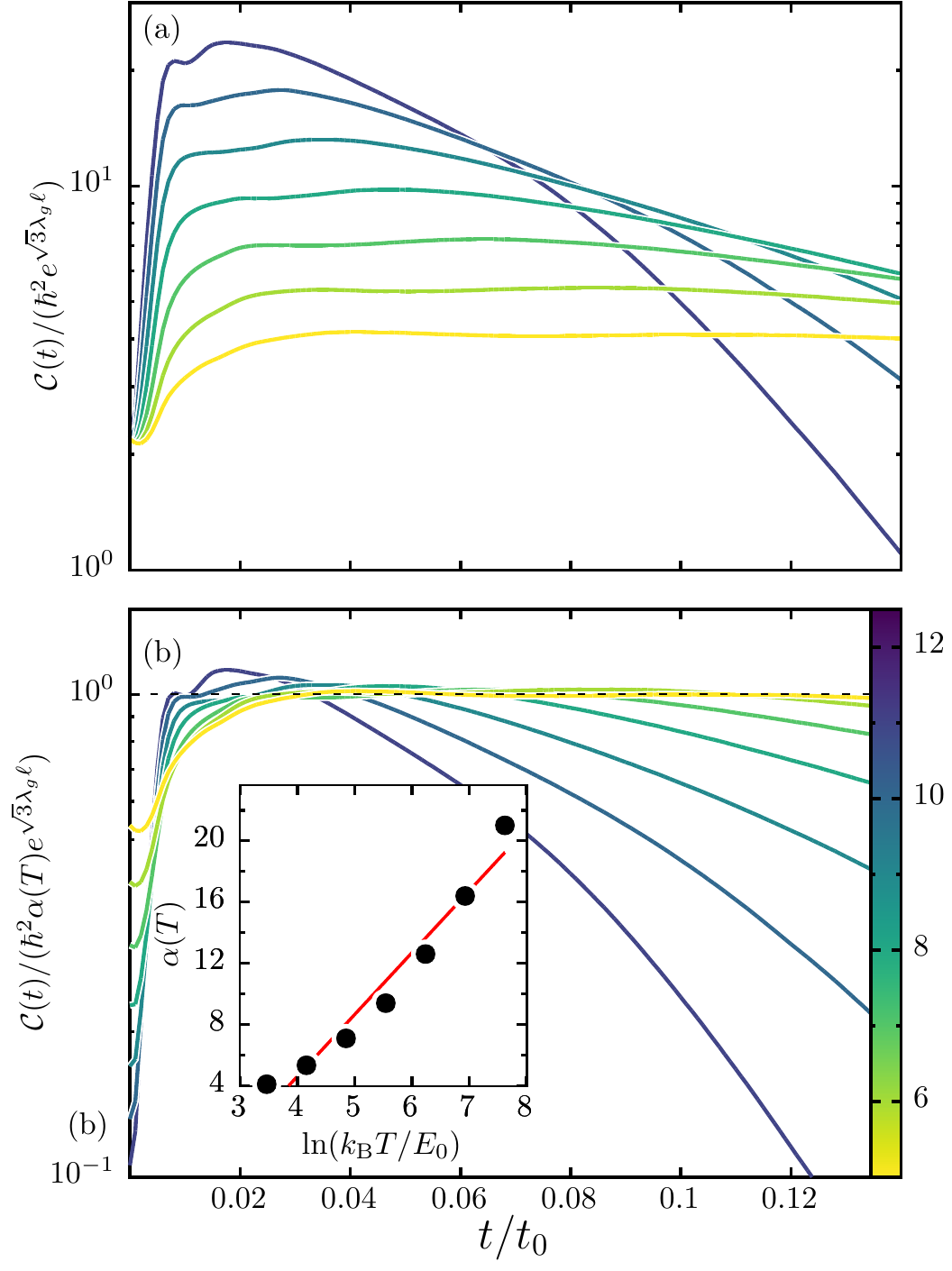}}
\caption{Same data as in Fig.~\ref{fig:tse}, showing the OTOC, scaled with $e^{\sqrt{3} \lambda_{\rm g} \ell}$ (a), and with $\alpha(T) e^{\sqrt{3} \lambda_{\rm g} \ell}$ (b), as a function of the time $t$  (scaled with $t_0=ma^2/\hbar$) for different temperatures according to the color scale (following the same conventions than in Fig.~\ref{fig:Oj}). The time-window where the exponential growth provides a good description of the OTOC increases upon decreasing the temperature. Inset: scaling parameter $\alpha(T)$ exhibiting an approximate logarithmic temperature dependence.
\label{fig:sr}}
\end{figure}

While the exponential fit performed in the interval $0.4 \lesssim \ell/a \lesssim 1.5$ is hindered by the limited range of growth, the rate predicted by Eq.~\eqref{eq:CLowT} provides a good description at all working temperatures. Moreover, for these values of $\ell/a$, it is meaningful to confront the quantum numerical results with the semiclassical calculations. We remark that trajectories longer than those with the mean-trajectory-length are included in the semiclassical expressions like Eq.~\eqref{eq:Csemiclass} and may  result in an important contribution. The exponential regime is further put in evidence in Fig.~\ref{fig:sr}(a), where we scaled ${\cal C}(t)$ with $\exp{(\sqrt{3} \lambda_{\rm g} \ell)}$. The time-window of the exponential growth increases when lowering the temperatures, in agreement with the theoretical prediction. A numerical prefactor $\alpha(T)$, with a weak (logarithmic) temperature dependence, characterizes this regime, [Fig.~\ref{fig:sr}(b)]. The fitting to the exponential form becomes poorer upon increasing the temperature, since, as discussed before Eq.~\eqref{eq:CLowT}, the thermal average does not simply select a mean velocity.  

The saturation of the OTOC for long times is expected for a finite-size system 
\cite {kgp17,cdp17,rur18} and has been observed in the numerical simulations performed in several physical cases \cite{rgg17,hmy17,letal17,gsjrw18}. From the inset of Fig.~\ref{fig:tse}, we see that the value of the long-time saturation of the OTOC is proportional to the temperature. In addition to the term given by Eq.~\eqref{eq:CLongT} (proportional to temperature), we should consider the contribution arising from the different long-time value of the OTOC components, discussed in Sec.~\ref{sec:QNCOTOCC}
(also proportional to temperature). Moreover, it is important to remark that quantifying the saturation values of quantities like the OTOC or the Loschmidt echo, whose semiclassical expression depend on four trajectories is difficult, due to the different possible pairings \cite{gg09} and the effect of trajectory loops \cite{sieber2001,gwgkr}. 

As in the case of Eq.~\eqref{eq:O3class}, the choice of the operator pair does not alter the structure of Eq.~\eqref{eq:Csemiclass}, but it merely modifies the phase-space variables within the curly bracket. These changes have important consequences for the final results, since in the thermal average the Boltzmann factor weights the states according to their energy. For instance, taking $A=B=X$ simply requires trading the term ${\hat P}_X^2$ inside the curly bracket by $X^2$. The stationary-phase procedure that leads to Eq.~\eqref{eq:CLowT}, now requires much lower temperatures to be applicable. The quantum numerical results for the OTOC in this case (not shown) exhibit a monotonous increase with time, up to a temperature-independent saturation value, but a window of exponential growth cannot be identified when the range of temperatures of Fig.~\ref{fig:tse} is investigated. Consistent with the previous analysis, the temperature dependence of the OTOC in the case $A=B=X$ is inverted with respect to that of the case $A=X$, $B=P_X$ of Fig.~\ref{fig:tse}. Taking $A=B=P_X$ results in very different OTOC components, with respect to the other two previously discussed cases, while for the associated OTOC a time window of exponential growth can be identified (not shown). 
\section{Conclusions}
\label{sec:conclusions}
The OTOC of quantum operators in low-dimensional systems has been addressed by semiclassical and numerical techniques. The four-point OTOC ${\cal C}(t)$ can be expressed in terms of three components ${\cal O}^{(j)}(t)$, for $j=1,2,3$. Introducing the semiclassical expression of the propagator, ${\cal O}^{(3)}(t)$, results as a sum of terms depending on pairs of trajectories, while the other two components, ${\cal O}^{(1)}(t)$ and ${\cal O}^{(2)}(t)$, require the simultaneous treatment of four classical trajectories. The leading-order $\hbar^0$ approximation for each of the three components of the OTOC is the same, such that at the classical level the OTOC vanishes for all times. The OTOC, stemming from the difference of these components is of order $\hbar^2$, and is much smaller than each of them.

The case of the position and momentum operators in a two-dimensional classically chaotic system has been chosen to illustrate the capability of semiclassical expansions to obtain the first leading order terms in the $\hbar$ expansion of the OTOC. Numerical quantum calculations of the OTOC and its components, using a unsymmetrical billiard, were carried out to test the semiclassical results and set their limit of validity. 

The leading-order (classical) approximation to the OTOC components was shown to provide a good description of the numerical results, and the differences detected for long-times were attributed to quantum corrections. The semiclassical approach to the OTOC yielded an exponential increase within a limited time-window between the initial fast increase and the saturation for large times. The corresponding growth rate was in good agreement with that of the numerical calculations. The semiclassical approach, as well as the numerical one, predict that the time-window exhibiting exponential growth of the OTOC becomes larger as the temperature of the system diminishes.

The bound on the OTOC growth-rate proposed in Ref.~\cite{mss16} is 
$\Lambda \le 4\pi \kb T/\hbar$, while we obtained 
$\Lambda=\sqrt{3} \ \lambda_{\rm g} {\tilde v}$. Thus, for the system and the operators considered, the proposed bound would hold if
$ \kb T \ge 3\hbar^2 \lambda_{\rm g}^2/(16 \pi^2 m)$. We then conclude that the fulfillment of the bound is guaranteed, unless $\kb T$ is of the order of the ground state of the billiard. Notwithstanding, the prediction of a global growth for the OTOC encounters two limitations. Firstly, the thermal average necessarily mixes different growths given by an energy-dependent Lyapunov exponent. Secondly, the finite-size of the systems under consideration necessarily leads to a saturation of the OTOC for large times.  

A different choice of operators results in similar semiclassical expressions to the ones obtained in this work, but the thermal average gives to the OTOC and its components a temperature dependence that is very sensitive to the nature of these operators. The choice of an initial thermal state results in the mixing of various growth rates, and therefore in a different behavior of the OTOC, with respect to the case of an initially localized wave-packet \cite{rgg18,hirschetal18,rfu18}, where such a mixing is not present.
 
The connection between the OTOC and the Loschmidt echo has been signaled in various contexts \cite{swingle18,zangaraphiltrans,k18,rfu18} as both link irreversibility with the operator noncommutativity. At the technical level, both observables are four-point functions that in a semiclassical formulation depend on four classical trajectories. Unlike the Loschmidt echo, there is only one Hamiltonian to be considered in the case of the OTOC, and this feature implies to take into account subtle action differences between nearby trajectories. Like in the calculation of the Loschmidt echo, the simplest semiclassical approach to the OTOC can be done for uniformly hyperbolic systems, and this has been the path followed in this work.  When the Loschmidt echo is averaged over different intial wave-packets, the local variations of the Lyapunov exponent encountered in a typical chaotic system lead to the hyper-sensibility with respect to the perturbation, where the average decay is governed by a modified exponent and is representative of typical conditions only after the Ehrenfest time  \cite{stb03}. The effects that emerge from the failure of reaching complete self-averaging of the Lyapunov exponent within a limited time-interval are more prominent in the case of the OTOC \cite{rgg18},  due to its initial exponential increase with time. 

\acknowledgments
We acknowledge helpful discussions with H.M. Pastawski, A.J. Roncaglia, and M. Saraceno. We thank M. Saraceno and D. Weinmann for a careful reading of the manuscript and valuable suggestions. The authors received funding from French National Research Agency (Project No. ANR- 14-CE36-0007-01), the CONICET (Grant No. PIP 11220150100493CO), the ANPCyT (PICT- 2016-1056), the UBACyT (Grant No 20020130100406BA), a bi-national collaboration project funded by CONICET and CNRS (PICS No. 06687), and
Laboratoire International Associ\'e (LIA) ``LiCOQ''.

\appendix 
\section{Alternative pairings}
\label{app:AP}

\begin{figure}
\centerline{\includegraphics[width=0.45\textwidth]{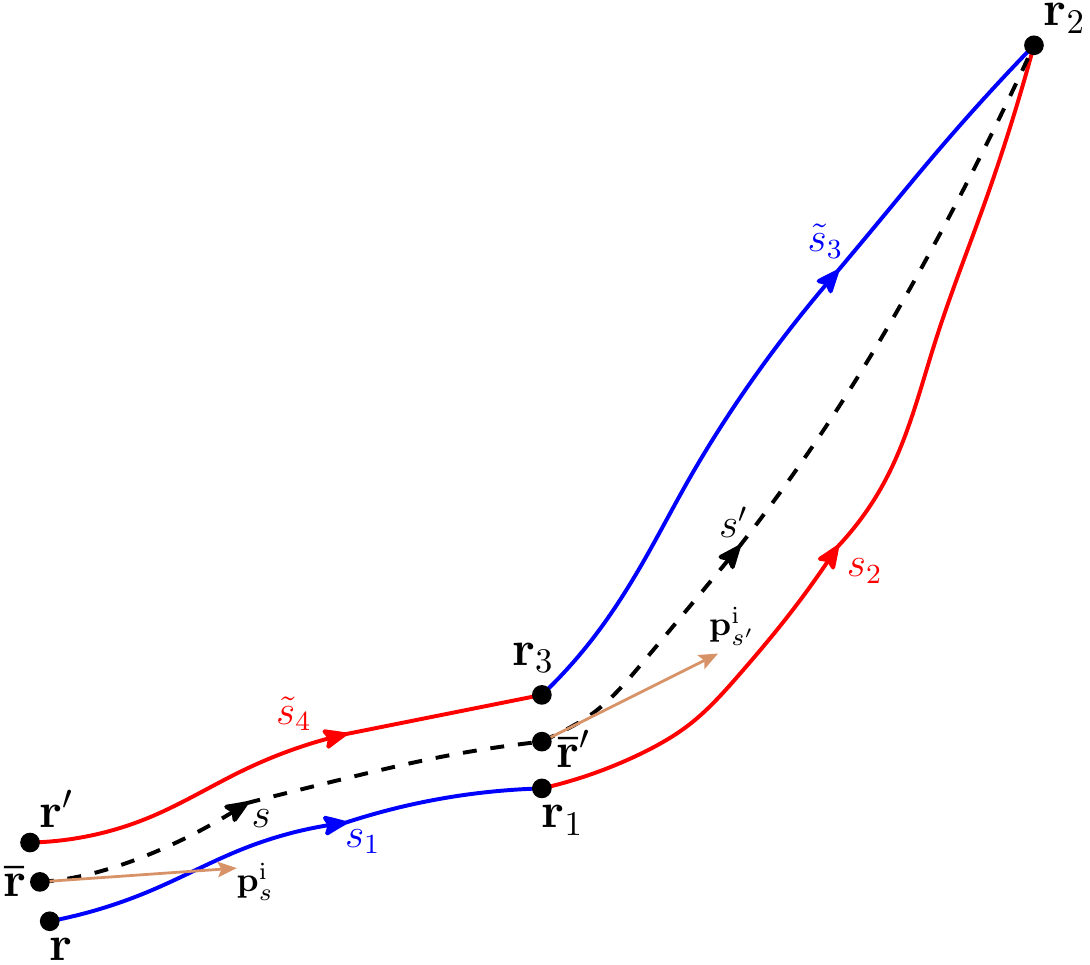}} 
\caption{Graphical representation of the $O^{(2)}_{\rm ap}(\rvp,\rv;t)$ term assumed for writing Eq.~\eqref{eq:O2ap}. The color blue (red) is used for trajectories whose Hamilton principal function appears with a plus (minus) sign in the phase term of Eq.~\eqref{eq:O2ap}. The dashed (black) lines represent the trajectories $s$ and $s^{\prime}$ (with initial momenta $\pv_{s}^{\rm i}$ and $\pv_{s^{\prime}}^{\rm i}$ respectively) used to linearize the dynamics of nearby trajectories.
\label{fig:O4}}
\end{figure}

As indicated in Sec.~\ref{sec:DSOTOCc}, the terms in the semiclassical expansion Eq.~\eqref{eq:O2semi} vanish unless their associated trajectory actions are somehow related. While the diagonal scheme provides the most obvious relationship, other possibilities might result in terms that survive the spatial integrations over the internal variables $\rv_1$, $\rv_2$, and $\rv_3$. Among them, that of Fig.~\ref{fig:O4}, where the trajectory $s_1$ ($s_2$) remains close to ${\tilde s}_4$ (${\tilde s}_3$). Within such {\it alternative pairing} (noted as ``ap''), performing the changes of variables $\rb=(\rv+\rvp)/2$, $\deltarv=\rvp-\rv$, $\rbp=(\rv_1+\rv_3)/2$ and $\deltarvp=\rv_3-\rv_1$, and proceeding in the same way that  lead to Eq.~\eqref{eq:O2diag}, we write
\begin{widetext}
\begin{align}
{\cal O}^{(2)}_{\rm ap}(t) &  = \frac{1}{\left(2\pi \hbar\right)^4 {\cal A}}  \int \d \rbp \ \d \deltarvp \ \d \rb \ \d \deltarv \  \d \rv_{2}  \
\exp{\left[-\frac{m}{2 \beta \hbar^2} \deltarv^2 \right]} \
\sum_{s(\rb,\rbp;t)} \ \sum_{s^{\prime}(\rbp,\rv_{2};t)} 
C_{s} \ C_{s^{\prime}} 
\nonumber \\ &
\left\{\left(\left({\bar X}^{\prime}\right)^2 - \frac{1}{4} \left(\deltarvp_{X}\right)^{2} \right)\left(P_{X,s_{3}}^{\rm i}  \right)^{2} \right\} \
\exp{\left[  \frac{\im}{\hbar} \left( \pv_{s}^{\rm i}.\deltarv - \pv_{s}^{\rm f}.\deltarvp\right) \right]  } \
\exp{\left[ - \frac{\im}{\hbar}  \pv_{s^{\prime}}^{\rm i}.\deltarvp \right]  } \, .
\label{eq:O2ap}
\end{align}
\end{widetext}
Where we have used that
\begin{subequations}
\label{allap} 
\begin{equation}
R_{s_{1}}(\rv_{1},\rv;t)-R_{{\tilde s}_{4}}(\rv_{3},\rvp;t) \simeq \pv_{s}^{\rm i}.\deltarv - \pv_{s}^{\rm f}.\deltarvp \, ,
\end{equation}
\begin{equation}
R_{{\tilde s}_{3}}(\rv_{2},\rv_{3};t)-R_{s_{2}}(\rv_{2},\rv_{1};t) \simeq - \pv_{s^{\prime}}^{\rm i}.\deltarvp  \, ,
\end{equation}
\end{subequations}
which are valid up to quadratic order in $\deltarv$ and $\deltarvp$. With the new integration variables, the trajectory $s_{3}$ is the one that leaves $\rv_{2}$ and reaches $\rbp + \deltarvp/2$ in time $t$, while remaining in the environment of ${\tilde s}^{\prime}$. 

In leading order we use a strict diagonal condition $\tilde{s}_{4} = s_{1}$, $\tilde{s}_{3} = s_{2}$ for the terms within the curly bracket. 
Making the change of variables from $\rv_{2}$ to $\pv_{s^{\prime}}^{\rm i}$ (that we note $\pvp$) and from $\rbp$ to $\pv_{s}^{\rm i}$ (that we note $\pv$), we perform the integral over $\deltarvp$, that results in a term $\delta \left(\pvp+\pv_{s}^{\rm f} \right)$. The delta function settles $\pvp$ to $-\pv_{s}^{\rm f}$, implying that $s^{\prime}$ is indeed oriented opposite than in the sketch of Fig.~\ref{fig:O4}. The resulting configuration is then that of Fig.~\ref{fig:O2}, showing that the alternative pairing does not produce in leading order a new contribution beyond the one of the diagonal scheme already considered.

\section{Periodic-orbit corrections}
\label{app:B}

Within the diagonal scheme, in Sec.~\ref{sec:DSOTOCc} we used the approximation Eq.~\eqref{eq:freeGF}, of only keeping the free-space form of the Green function. To go beyond such an approximation, we can use the semiclassical form of the Green function \cite{LesHou89, Gutz-book}
\begin{align}
G_{\rm sc}(\rvp,\rv;\varepsilon) &= \frac{1}{\sqrt{2\pi} \hbar^{3/2}}  \sum_{s(\rv,\rvp;\varepsilon)} D_{s}^{1/2} 
\nonumber \\
&\exp{\left[\tfrac{\im}{\hbar}S_{s}(\rvp,\rv;\varepsilon)-\im \tfrac{\pi}{2}\nu_{s}-\im \tfrac{3\pi}{4}\right]  }
\  \, ,
\label{eq:SCGF}
\end{align}
obtained as the Fourier transform of the semiclassical propagator Eq.~\eqref{eq:SCpropagator}. The sum is over all the classical trajectories $s(\rv,\rvp;\varepsilon)$ joining the points $\rv$ and $\rvp$, with an energy  $\varepsilon$, an action $S_{s}(\rvp,\rv;\varepsilon)$, and a number of conjugate points $\nu$. The factor $D_{s}$ is given in terms of second derivatives of the action with respect to the energy $\varepsilon$ and the initial and final space coordinates $\rv$, $\rvp$. 

\begin{widetext}
The component ${\cal O}^{(3)}(t)$ can then be written as
\begin{equation}
{\cal O}^{(3)}(t) = \frac{1}{2\im} \left\{ {\cal O}^{(3)}_{+}(t) - {\cal O}^{(3)}_{-}(t) \right\}
\  \, ,
\label{eq:O3PO1}
\end{equation}
with
\begin{align}
{\cal O}^{(3)}_{+}(t)   &  = 
\frac{2}{Z} \frac{1}{(2 \pi \hbar)^{7/2}} \int  \d \ve \ \d \rvp \ \d \rv_{1} \  \d \rv \ e^{-\beta \ve} \ 
\sum_{s_{2}(\rv_{1},\rvp;t)} \ \sum_{s_{1}(\rv,\rv_{1};t)} 
C_{s_{2}}^{1/2} \ C_{s_{1}}^{1/2} \
\left\{P_{X,s_{2}}^{\rm f} \left( X_{1} \right)^{2} P_{X,s_{1}}^{\rm i}\right\} 
\nonumber \\ &
\exp{\left[  \frac{\im}{\hbar}\left(R_{s_{1}}(\rv_{1},\rv;t)-R_{s_{2}}(\rvp,\rv_{1};t)\right)  -\im \frac{\pi}{2}\left( \mu_{s_{1}}-\mu_{s_{2}}\right)  \right]  } 
\nonumber \\ &
\sum_{s_{3}(\rvp,\rv;\ve)} D_{s_{3}}^{1/2}  
\exp{\left[\tfrac{\im}{\hbar}S_{s_{3}}(\rv,\rvp;\varepsilon)-\im \tfrac{\pi}{2}\nu_{s_{3}}-\im \tfrac{3\pi}{4}\right]  }
\, ,
\label{eq:O3PO2}
\end{align}
\end{widetext}
The previous expression is particularly involved, as it mixes time and energy-dependent semiclassical expansions. Temperature-dependent semiclassical approaches have been developed for the stationary case \cite{ullmo97,ullmo98}, but the time-dependence of Eq.~\eqref{eq:O3PO2} forces us to take another route. 

Within the semiclassical approach for completely chaotic systems, it is natural to evaluate the spatial integrals in Eq.~\eqref{eq:O3PO2} over the rapidly oscillating integrand by the stationary-phase method. The stationary-phase conditions for the $\rv$, $\rv_{1}$, and $ \rvp$ integrations give, respectively, 
\begin{subequations}
\label{allSPC} 
\begin{equation}
\pv_{s_{1}}^{\rm f}(\rv_{1},\rv;t) + \pv_{s_{2}}^{\rm i}(\rvp,\rv_{1};t) =  0  \ ,
\label{eq:SPCa}
\end{equation}
\begin{equation}
-\pv_{s_{1}}^{\rm i}(\rv_{1},\rv;t) + \pv_{s_{3}}^{\rm f}(\rv,\rvp;\ve) =  0   \ ,
\label{eq:SPCb}
\end{equation}
\begin{equation}
-\pv_{s_{2}}^{\rm f}(\rvp,\rv_{1};t) - \pv_{s_{3}}^{\rm i}(\rv,\rvp;\ve) =  0   \ .
\label{eq:SPCc}
\end{equation}
\end{subequations}
In principle, trajectories $s_{1}$ and $s_{2}$ are not traveled with the same energy, but the condition Eq.~\eqref{eq:SPCa} imposes such a restriction. Similarly, the energy with which $s_{1}$ and $s_{2}$ are traveled is, in principle, independent of $\ve$, but the conditions Eqs.~\eqref{eq:SPCb} and \eqref{eq:SPCc} impose the previous condition. 

\begin{figure}
\centerline{\includegraphics[width=0.48\textwidth]{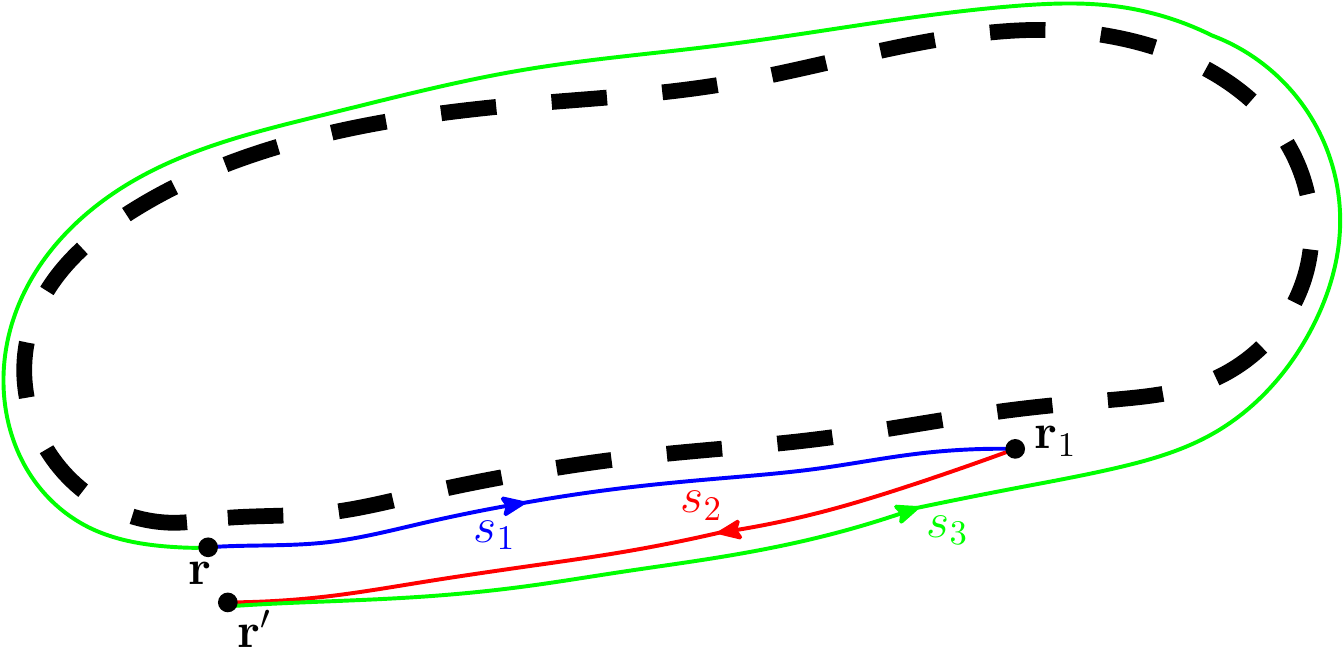}} 
\caption{Graphical representation of  ${\cal O}^{(3)}_{+}(t)$ according to Eq.~\eqref{eq:O3PO2} in the case where the trajectories $s_{1}$, $s_{2}$, $s_{3}$ remain close to a periodic orbit (thick dashed). The color blue (red) is used for fixed-time trajectories whose Hamilton principal function appears with a plus (minus) sign in the phase term of Eq.~\eqref{eq:O3PO2}, while green is reserved for fixed-energy trajectories. 
\label{fig:O5}}
\end{figure}

The most obvious solution of Eqs.~\eqref{allSPC} results in the scheme worked out for ${\cal O}^{(3)}(t)$ in Secs.~\ref{sec:DSOTOCc} and \ref{sec:LOCOTOCC}. However, there is an additional possibility of satisfying the conditions Eqs.~\eqref{allSPC} when the three stationary-phase points in consideration are part of a classical periodic orbit (see Fig.~\ref{fig:O5}). A similar configuration has been shown to provide the periodic-orbit corrections to the Drude conductivity in the phase coherent case \cite{richter95,hacken95,hacken95b}. The stationary-phase condition for ${\cal O}^{(3)}_{-}(t)$ results in the same configuration of Fig.~\ref{fig:O5}, with the only modification of inverting the sens in which the trajectory $s_{3}$ is traveled. 

The calculation of ${\cal O}^{(3)}_{+}(t)$ proceeds in following the footprints of Ref.~[\onlinecite{hacken95b}]. The additional difficulty with respect to the previous case it that we need to handle now three spatial integrations instead of two. We thus highlight the steps of the calculation arising from this feature. Introducing a new set of orthogonal coordinates along the periodic trajectory in the parallel ($\parallel$) and in the perpendicular ($\perp$) directions, {\it i.e.} $\rv=(r^{\parallel },r^{\bot})$, $\rvp=(r^{\prime \parallel },r^{\prime \bot})$, $\rv_{1}=(r_{1}^{\parallel },r_{1}^{\bot})$, the integration over the transverse coordinates is given in terms of the determinant of the quadratic form associated with 
\begin{equation}
Q = \left(
\begin{array}{ccccc}
 \frac{\partial^{2}R_{s_{1}}}{\partial r^{\bot 2}} + \frac{\partial^{2}S_{s_{3}}}{\partial r^{\bot 2}} 
 & \hspace{0.cm} & \frac{\partial^{2}S_{s_{3}}}{\partial r^{\bot}\partial r^{\prime \bot}} 
 & \hspace{0.cm} & \frac{\partial^{2}R_{s_{1}}}{\partial r^{\bot}\partial r_{1}^{\bot}}  \\
 \frac{\partial^{2}S_{s_{3}}}{\partial r^{\bot}\partial r^{\prime \bot}} 
 & \hspace{0.cm} & - \frac{\partial^{2}R_{s_{2}}}{\partial r^{\prime \bot 2}} + \frac{\partial^{2}S_{s_{3}}}{\partial r^{\prime \bot 2}} 
& \hspace{0.cm} & - \frac{\partial^{2}R_{s_{2}}}{\partial r^{\prime \bot}\partial r_{1}^{\bot}}   \\
\frac{\partial^{2}R_{s_{1}}}{\partial r^{\bot}\partial r_{1}^{\bot}} 
& \hspace{0.cm} & - \frac{\partial^{2}R_{s_{2}}}{\partial r^{\prime \bot}\partial r_{1}^{\bot}} 
& \hspace{0.cm} & \frac{\partial^{2}R_{s_{1}}}{\partial r_{1}^{\bot 2}} - \frac{\partial^{2}R_{s_{2}}}{\partial r_{1}^{\bot 2}}  
\end{array}
\right) 
\end{equation}
The matrix elements of $Q$ can be expressed in terms of those of the monodromy matrix $M$ of the selected periodic orbit \cite{hacken95b}.

The integration over the longitudinal coordinates can be easily done in the case of a billiard and leads to 
\begin{align}
{\cal O}^{(3)}_{\rm PO}(t)  & = \frac{2}{(2 \pi \hbar)^{2}} \int  \d \ve  \ e^{-\beta \ve} \nonumber \\ 
&
\sum_{\gamma} \ \sum_{p=1}^{\infty} c_{\gamma}(t) \ 
\frac{1}{|{\rm Tr}M_{\gamma}^{p}-2|^{1/2}} \ \cos{\left[\tfrac{p}{\hbar}S_{\gamma}(\varepsilon)-\tfrac{p \pi}{2}\nu_{\gamma}\right]  } \, ,
\label{eq:O3PO3}
\end{align}
as a sum over the primitive periodic orbits $\gamma$ and their repetitions $p$. We have defined the $P_{X}\!-\!X$ time-dependent correlation function along the length $L_{\gamma}$ of the primitive periodic orbit as
\begin{equation}
c_{\gamma}(t) =  \int_{0}^{L_{\gamma}}  \d l \ \{P_{X}^{2}(l) \ X^{2}(l;t)\} 
 \, .
\label{eq:cf}
\end{equation}
 
Equation~\eqref{eq:O3PO3} has the same structure of the trace formula for the density of states \cite{LesHou89,Gutz-book}, as well as of the phase-coherent conductivity \cite{richter95,hacken95,hacken95b} and the response function of a many-body fermionic system \cite{gaspard_jain97}, differing from them only in the particular correlation function and the thermal averaging. Like in the well-studied previous cases, the actual evaluation of \eqref{eq:O3PO3} is in general quite difficult, since it requires the determination of a large number of periodic orbits with their associated parameters. Therefore, its use is more formal than practical. 
Similarly as in Sec.~\ref{sec:LOCOTOCC}, we remark that the result Eq.~\eqref{eq:O3PO3} directly follows from the periodic-orbit correction to the thermal average of the operators explicit in  Eq.~\eqref{allO}. The curly bracket appearing in the definition of the correlation function is the same as in Eq.~\eqref{eq:O3class}. The difference between them being that in the previous case the integration is over the area ${\cal A}$ of the system, while in Eq.~\eqref{eq:cf} the integration carries over the initial point on the primitive periodic orbit $\gamma$ (parametrized by $l$), while $X(l;t)$ is given by the evolution on the periodic orbit and $P_X$ is the $X$-component of the initial momentum . 

In the leading-order in $\hbar$, the periodic-orbit correction ${\cal O}^{(3)}_{\rm PO}(t)$ is the same that for the other two OTOC components. Therefore, such  corrections do not affect ${\cal C}(t)$. However, as discussed in Sec.~\ref{sec:QNCOTOCC}, they can be important in determining the long-time saturation value and the oscillations of OTOC components.


\begin{thebibliography}{99}

\bibitem {swingle18}
B. Swingle; Nature Physics \textbf{14}, 988 (2018).

\bibitem {lo69}
A. Larkin and Y.N. Ovchinnikov; Zh. Eksp. Teor. Fiz. Sov. Phys. \textbf{55}, 1200 (1968); JETP \textbf{28}, 1200 (1969).

\bibitem {mss16}
J. Maldacena, S H. Shenker, and D. Stanford; J. High Energy Phys. \textbf{08}, 106 (2016).

\bibitem {sbsh16}
B. Swingle, G. Bentsen, M. Schleier-Smith, and P. Hayden;
Phys. Rev. A \textbf{94}, 040302(R) (2016).

\bibitem {afi16}
I.L. Aleiner, L. Faoro, and L.B. Ioffe; Annals Phys. \textbf{375}, 378 (2016). 

\bibitem {cg16}
M. Campisi and J. Goold; Phys. Rev. E \textbf{95}, 062127 (2017).

\bibitem {kgp17} 
I. Kukuljan, S. Grozdanov, and T. Prosen;
Phys. Rev. B \textbf{96}, 060301(R) (2017).

\bibitem {rgg17}
E.B. Rozenbaum, S. Ganeshan, and V. Galitski; Phys. Rev. Lett.  \textbf{118}, 086801 (2017).

\bibitem {hmy17}
K. Hashimoto, K. Muratab, and R. Yoshii; J. High Energy Phys. \textbf{10}, 138 (2017). 

\bibitem {letal17} 
J. Li {\it et al}; Phys. Rev. X \textbf{7}, 031011 (2017).

\bibitem {cdp17}
J.S. Cotler, D. Ding, and G.R. Penington; arXiv:1704.02979.

\bibitem {gbswb17}
M. G\"arttner, J.G. Bohnet, A. Safavi-Naini, M.L. Wall, J.J. Bollinger, and A.M. Rey; Nature Physics \textbf{13}, 781 (2017).

\bibitem {k18}
J. Kurchan; J. Stat. Phys. \textbf{171}, 965 (2018).

\bibitem {rgg18}
E.B. Rozenbaum, S. Ganeshan, and V. Galitski; arXiv:1801.10591 (2018).

\bibitem {cz18}
X. Chen and T. Zhou; arXiv:1804.08655 (2018).

\bibitem {rur18}
J. Rammensee, J. D. Urbina, and K. Richter;
Phys. Rev. Lett., in press (2018).

\bibitem {gsjrw18}
I. Garc\'{\i }a-Mata, M. Saraceno, R.A. Jalabert, A.J. Roncaglia, and D.A. Wisniacki;
Phys. Rev. Lett. {\bf 121} 210601 (2018).

\bibitem {rfu18}
R. Hamazaki, K. Fujimoto, and M. Ueda; arXiv:1807.02360 (2018).

\bibitem {hirschetal18}
J. Ch\'avez-Carlos, {\it et al.}; arXiv:1807.10292 (2018).

\bibitem {bgs84}
O. Bohigas, M.J. Giannoni, and C. Schmit; Phys. Rev. Lett., \textbf{52}, 1 (1984).

\bibitem {ullmo16}
D. Ullmo; Scholarpedia; 11(9):31721 (2016).

\bibitem {LesHou89}M.-J.~Giannoni, A.~Voros, and J.~Zinn-Justin, eds.,
\textit{Chaos and Quantum Physics }(North-Holland, Amsterdam, 1991).

\bibitem {Gutz-book}M.C.~Gutzwiller, \textit{Chaos in Classical and Quantum
Mechanics} (Springer-Verlag, Berlin, 1990); and in Ref. \cite{LesHou89}.

\bibitem {jalabert2001} R.A.~Jalabert and H.M.~Pastawski; Phys. Rev.
Lett. \textbf{86}, 2490 (2001).

\bibitem{goussev12}
A. Goussev, R.A. Jalabert, H.M. Pastawski, D.A. Wisniacki;  Scholarpedia, 7(8):11687 (2012).

\bibitem{philtrans}
A. Goussev, R.A. Jalabert, H.M. Pastawski, D.A. Wisniacki, eds.,
\textit{Loschmidt echo and time reversal in complex systems};
Phil. Trans. R. Soc. {\bf A 374}, 20150383 (2016).

\bibitem{srednicki94}
M. Srednicki; Phys. Rev. E {\bf 50}, 888 (1994).

\bibitem{alonso_jain97}
D. Alonso and S.R. Jain; 
J. Phys. A {\bf 30}, 4993 (1997).

\bibitem{deutsch18}
J.M. Deutsch;
Rep. Progr. Phys.  {\bf 81}, 082001 (2018).

\bibitem{dima1}
D.L.Shepelyanskii;
Theor. Math. Phys. {\bf 49}, 925 (1981).

\bibitem{dima2}
D.L.Shepelyansky;
Physica {\bf 8D}, 208 (1983).

\bibitem{GR}
See expression 6.631.4 in I.S. Gradshteyn and I.M. Ryzhik,
\textit{Table of integrals, series and products} (Academic Press, San Diego, 1980).

\bibitem{vergini95} 
E. Vergini and M. Saraceno; Phys. Rev.  E {\bf 52}, 2204 (1995).

\bibitem {Goussev2008} A. Goussev, D. Waltner, K. Richter, and R.A. Jalabert; New Journal of Physics {\bf 10} 093010 (2008).

\bibitem{ullmo97} D. Ullmo, K. Richter, H.U. Baranger, F. von Oppen, and R.A. Jalabert; Physica E {\bf 1}, 268 (1997).

\bibitem{ullmo98} D. Ullmo, H.U. Baranger, K. Richter, F. von Oppen, and R.A. Jalabert; Phys. Rev. Lett. {\bf 80}, 895 (1998).

\bibitem {richter95}
K. Richter; Europhys. Lett., {\bf 29}, 7 (1995).

\bibitem{hacken95} 
G. Hackenbroich and F. von Oppen; Europhys. Lett., {\bf 29}, 151 (1995).

\bibitem{hacken95b} 
G. Hackenbroich and F. von Oppen; Z. Phys. B {\bf 97}, 157 (1995).

\bibitem{gg09}
M. Guti\'errez and A. Goussev; Phys. Rev. E {\bf 79}, 046211 (2009).

\bibitem{sieber2001} M. Sieber and K. Richter;  Physica Scripta {\bf T90}, 128 (2001).

\bibitem{gwgkr}
B. Gutkin, D. Waltner, M. Guti\'errez, J. Kuipers, and K. Richter;
Phys. Rev. E {\bf 81}, 036222 (2010).

\bibitem{zangaraphiltrans}
P.R. Zangara, D. Bendersky, P.R. Levstein, and H.M. Pastawski, in Ref.~\cite{philtrans}; Phil. Trans. R. Soc. {\bf A 374}, 20150163 (2016). 

\bibitem{stb03}
P.G. Silvestrov, J. Tworzyd\l o, and C.W.J. Beenakker;
Phys. Rev.  E {\bf 67}, 025204(R) (2003).

\bibitem{gaspard_jain97}
P. Gaspard and S. R. Jain;
Pramana {\bf 48}, 503 (1997).

\end{thebibliography}
\end{document}